\documentclass[11pt]{article}
\usepackage{graphicx}
\usepackage{subfigure,amssymb,amsmath}
\usepackage{cite,color,url}
\usepackage[colorlinks=true
,urlcolor=blue
,anchorcolor=blue
,citecolor=blue
,filecolor=blue
,linkcolor=blue
,menucolor=blue
,pagecolor=blue
,linktocpage=true
,pdfproducer=medialab
]{hyperref}

\renewcommand{\baselinestretch}{1.3}                                
\newcommand{\comment}[1]{}
\newcommand{\newc}{\newcommand}

\def\issue(#1,#2,#3){{\bf #1}, #2 (#3)}

\def\PREP(#1,#2,#3){Phys.\ Rep. \issue(#1,#2,#3)}
\def\EPJC(#1,#2,#3){Eur.\ Phys.\ J.\ C \issue(#1,#2,#3)}

\def\bul{$\bullet$ }

\def\r2{\sqrt 2}
\def\beq{\begin{equation}}
\def\eeq{\end{equation}}
\def\beqn{\begin{eqnarray}}
\def\eeqn{\end{eqnarray}}
\def\sinW2{\sin^2\theta_W}

\def\mz2{M_{z}^2}
\def\c2b{\cos 2\beta}

\def\m#1{{\tilde m}_#1}

\def\mz{M_Z}
\def\m0{m_0}

\def\rmGeV{\rm ~GeV}
\def\rmTeV{\rm ~TeV}

\def\sec2w{sec^2\theta_W}

\def\tanbeta{{\rm tan}\beta}
\def\gmin2{(g-2)_\mu}

\catcode`\@=11 

\def\lsim{\mathrel{\mathpalette\@versim<}}
\def\gsim{\mathrel{\mathpalette\@versim>}}
\def\@versim#1#2{\vcenter{\offinterlineskip
    \ialign{$\m@th#1\hfil##\hfil$\crcr#2\crcr\sim\crcr } }}

\newc{\wt}{\widetilde}
\newc{\ra}{\rightarrow}
\newc{\s}{\smallskip}
\newc{\nn}{\noindent}
\newc{\non}{\nonumber}

\def \chonep{{\wt\chi_1}^{+}}
\def \chonem{{\wt\chi_1^-}}
\def \chonep2{{\wt\chi_2^+}}
\def \chonem2{{\wt\chi_2^-}}

\def \chonem {{\wt\chi_1^\pm}}
\def \chargino2 {{\wt\chi_2^\pm}}

\def \ch2m {{\wt\chi_2^-}}

\def \chonep {{\wt\chi_1^+}}

 \def\mygraphh#1#2{ \subfigure[]{
    \label{#1}
    \vspace*{-5.0cm}
    \hspace*{-0.5in}
    \begin{minipage}[b]{0.5\textwidth}
    \centering
    \hspace*{4ex}
     \includegraphics[width=1.0\textwidth,height=1.0\textwidth]{#2}
    \vspace*{-8ex}
    \end{minipage}}
    \vspace*{-1ex}}
\def\mygraph#1#2{ \subfigure[]{
   \label{#1}
   \hspace*{-0.6in}
   \begin{minipage}[b]{0.5\textwidth}
   \centering
   \hspace*{4ex}
   \includegraphics[width=0.95\textwidth,height=0.9\textwidth]{#2}
   \vspace*{-4ex}
   \end{minipage}}
   \vspace*{-1ex}
}

\def \PMET{p{\!\!\!/}_T}
\def \EMET{E{\!\!\!/}_T}
\DeclareMathAlphabet{\mathpzc}{OT1}{pzc}{m}{it}
\newcommand{\hsm}{{\large {$\mathpzc{h}$}}}
\begin{document}
\begin{flushright}
{IPMU13-0096}
\end{flushright}
\begin{center}
{\large \bf Implications of 98 GeV and 125 GeV Higgs scenario in 
non-decoupling SUSY with updated ATLAS, CMS and PLANCK data}
\vskip 0.3cm
Biplob Bhattacherjee$^{a}$\footnote{biplob.bhattacherjee@ipmu.jp,$^2$tpmc@iacs.res.in,
$^3$tpac@iacs.res.in,$^4$tpuc@iacs.res.in,$^5$debottam.das@physik.uni-wuerzburg.de,
$^6$tpdkg@iacs.res.in},
Manimala Chakraborti$^{b2}$,
Amit Chakraborty$^{b3}$,
Utpal Chattopadhyay$^{b4}$,
Debottam Das$^{c5}$, and
Dilip Kumar Ghosh$^{b6}$
\vskip 0.5cm
{$^a$ 
Kavli Institute for the Physics and Mathematics of the Universe (WPI),\\
 The University of Tokyo, Kashiwa, Chiba 277-8583, Japan
}\\
{$^b$  Department of Theoretical Physics, Indian Association 
for the Cultivation of Science,\\  
2A \& B Raja S.C. Mullick Road, Jadavpur, 
Kolkata 700 032, India}\\
{$^c$ 
Institut f{\"u}r Theoretische Physik und Astrophysik, Universit{\"a}t W{\"u}rzburg, \\
       Am Hubland, 97074 W{\"u}rzburg, Germany
}
\end{center}



\begin{abstract}
We discuss both MSSM and NMSSM scenarios in which the lightest Higgs boson
 with $m_h=98$~GeV 
is consistent with the small excess ($\sim 2.3 \sigma$) observed at the LEP in $e^+ e^-
\rightarrow Zh$, with $h \rightarrow  b {\bar b}$
process and the heavier Higgs boson of mass close to 125~GeV as
the observed candidate of
the SM Higgs like particle at the LHC.  We show the allowed regions in the
non-decoupling Higgs zone of 
MSSM parameter space which are consistent with several low energy constraints
coming from heavy flavour physics, latest experimental data on Higgs
signals and lower
limit on superparticle masses from 7~TeV and 8~TeV LHC run. We also
implement the constraints from the
relic density of the cold dark matter as obtained from the recent PLANCK
data.  Additionally, we 
discuss the possibility of observing the light Higgs boson of mass 98~GeV at the
14~TeV LHC run via $pp \rightarrow V h$, with $h \rightarrow b \bar b$ 
using the technique
of jet substructure. Our analysis
shows that at 14~TeV LHC run with 300 ${\rm fb}^{-1}$ luminosity the signal
efficiency of such a light Higgs boson
is at most 2.5$\sigma$. Finally, we 
make a comment on the prospect of
proposed $e^+ e^-$ ILC to discover/exclude
this light Higgs boson.
\end{abstract}
\noindent
\section{Introduction}
\label{introductionsection}
The Large Hadron Colllider (LHC) 
announcements on discovery of the Higgs 
boson (\hsm) like particle as analyzed by the 
CMS\cite{Chatrchyan:2012ufa} 
and ATLAS\cite{Aad:2012tfa} collaborations have been highly encouraging for 
particle physicists who had been in search of the Higgs boson  
through several decades. Additionally, CDF and D0 experiments of the 
Tevatron 
at Fermilab also announced a result on 
\hsm$\rightarrow b \bar b$ where Higgs boson is produced in the associated 
vector boson mode $W$\hsm/$Z$\hsm\cite{Aaltonen:2012qt}.  
Within the Standard Model (SM)\cite{sm} 
one can compute Higgs production cross sections 
via different means as well as consider decay of Higgs to various final 
products\cite{Djouadi:2005gi}. 
For the production mode, the main contributing sources are (i) gluon-gluon fusion (ggF),
(ii) vector boson fusion (VBF) and (iii) associated production with 
vector bosons (V\hsm), where V generically stands for W and Z-bosons. 
The decay channels 
which are directly concerned with detection and prediction 
of Higgs mass, are $Z Z^*$, $W W^*$, 
$b \bar b$, $\gamma \gamma$, $\tau^+ \tau^-$ etc. Here $V^*$ refers to 
off-shell vector boson indicating \hsm$\rightarrow V V^* \rightarrow V f 
\bar f$. The two photon $(\gamma \gamma )$ final state refers to loop 
induced processes involving W-boson and heavy fermions like 
$t, b$-quarks and $\tau $-lepton in the loop. The  
dominance of a heavier fermion rather than a lighter one  
is of course due to larger  \hsm$f \bar f$ coupling. 
The recent LHC observations of Higgs are associated with 
the following channels: 
$\gamma \gamma$, $Z Z^*$~$(\rightarrow 4 l)$, $W W^*$~$(\rightarrow l \nu 
l \nu,~{\rm or}~ \rightarrow l \nu j j)$.  LHC observations in 
the di-photon and the 4-lepton decay channels indeed lead to the discovery of 
Higgs boson with mass $\sim$125~GeV at around $5\sigma$ level  
\cite{cms_diphoton,atlas_diphoton}.  
The ongoing Higgs 
search at LHC is very important because it may successfully 
probe the origins of 
electroweak symmetry breaking and it may also provide valuable information 
for scenarios Beyond the Standard Model (BSM) of particle physics.  
Undoubtedly, 
low energy supersymmetry (SUSY) has been a promising direction 
to travel in order to 
look for BSM effects in particle physics. 
The modest $N=1$ supersymmetric extension of SM, namely the 
Minimal Supersymmetric Standard Model (MSSM)\cite{mssm} has two Higgs doublets and 
this leads to several Higgs scalars. The MSSM Higgs states are:  
two CP-even neutral 
Higgs scalar bosons $h$ and $H$, one CP-odd neutral Higgs scalar boson $A$, 
and two charged Higgs scalar bosons $H^\pm$ ~\cite{mssm,Djouadi:2005gj}.
Although MSSM has a large number of parameters, the 
Higgs sector, while not considering any radiative corrections can be 
described by only a few parameters.  Apart from the experimentally 
determined mass of Z-boson $M_Z$, these are:  
i) $M_A$, the pseudoscalar Higgs mass parameter and ii) $\tan\beta$, the ratio 
of the Higgs vacuum expectation values $<H_2>/<H_1>$, where 
$H_1$ and $H_2$ give masses to down type of quarks and leptons and 
up type of quarks respectively. 
However,  
there can be a significant amount of radiative corrections to the 
masses of the Higgs bosons and this causes 
a few other MSSM parameters to be relevant like the 
parameters that are involved in top squark and to some extent bottom squark or 
even tau-slepton 
masses\cite{Djouadi:2005gj,mssmhiggs}. MSSM has a nice prediction of a 135~GeV upper limit for 
the mass $m_h$ of the lightest Higgs boson\cite{mssmhiggsloop}. 
Finding the observed Higgs boson to have a mass in the vicinity of 
$125 $~GeV at LHC 
is very exciting for pursuing SUSY as a scenario beyond the SM.

\noindent
The MSSM Higgs sector can be broadly divided into a 
decoupling and a non-decoupling zone\cite{Djouadi:2005gj}. 
The former region corresponds to pseudoscalar mass $M_A$ becoming very large 
(in practice above 300 GeV or so). In this limit, $\cos^2(\beta-\alpha)$ 
becomes vanishingly small (consequently $\sin^2(\beta-\alpha) \rightarrow 1$).
Here, $\alpha$ refers to the Higgs mixing angle.  
The couplings like $W^+W^-H$, $HZZ$, $ZAh$, $W^\pm H^\mp h$, 
$Z W^\pm H^\mp h$, $\gamma W^\pm H^\mp h$ are all proportional to 
$\cos(\beta-\alpha)$. On the other hand, $W^+ W^- h$, $hZZ$, $ZAH$, $W^\mp H^\pm H$,
$Z W^\mp H^\pm H$  and $\gamma W^\mp H^\pm H$ couplings are all 
proportional to 
$\sin(\beta-\alpha)$. In the decoupling limit, an obvious
possibility is to interpret the newly observed state at around 125 GeV as the
light $CP$-even Higgs boson having SM-like couplings (for details see 
Refs.\cite{decouplingrefs,Carena:2013iba,Djouadi:2013vqa}).  
As a consequence, all other Higgs bosons become much 
heavier than $h$.  The non-decoupling 
region on the other hand is specified by $M_h \sim M_A \sim M_H \sim 
M_Z$ or $\sin^2(\beta-\alpha)$ becoming very small. 
In the LHC context, the non-decoupling region with $\cos^2(\beta-\alpha)$ 
becoming close to unity would mean larger coupling strength 
of $H$ with the electroweak 
gauge bosons. Thus here we may explore the possibility 
of $m_H \sim$~ 125 GeV, instead of considering the lighter counterpart $h$ to 
correspond to the discovered boson.  Here the heavier Higgs field $H$
would have SM-like couplings whereas the lighter 
Higgs $h$ would have much weaker couplings to W and Z-bosons, 
for details see Ref.\cite{Heinemeyer:2011aa,nondecouplingrefs,Bechtle:2012jw,Hagiwara:2012mga}.

We further consider $m_h$ to be in the vicinity of 98~GeV in the light of 
an old result from LEP Collaborations in regard to $e^+e^- \rightarrow Z$\hsm  
~with \hsm$\rightarrow b \bar b$ where the result 
indicated a possibility of seeing an  
excess of Higgs-like events\cite{Barate:2003sz} near the above value of 
mass. This excess came with a significance level of 2.3 times the 
standard deviations in the combined analysis of all the four LEP experiments. 
It could not be explained 
via an SM-like Higgs boson because of a low  
production cross section. 
Considering the discovered 
Higgs boson to be $H$ with $m_H\sim 125$~GeV rather than $h$, 
MSSM in principle, in its non-decoupling domain can accommodate 
$m_h$ at 98 GeV. Smaller $\sin^2(\beta-\alpha)$, consequently 
smaller $hZZ$ or $hWW$
couplings would be consistent with 
the LEP data whereas $H$ with its SM-like couplings to vector 
bosons would provide us with consistent $H \rightarrow \gamma \gamma$ 
and  $H \rightarrow Z Z^* \rightarrow 4l$, the four-lepton decay 
channels. In a pre-LHC period, such a possibility of having $m_h \sim 98$~GeV 
in the context of the LEP data was analyzed in 
Ref.\cite{Drees:2005jg}. In the recent time, in the 
``{\it post-Higgs at 125 GeV''} scenario the above 
Inclusive LEP-LHC Higgs perspective of assuming  
$m_h$ at 98 GeV along with $m_H$ at 125 GeV in the context 
of MSSM was analyzed in Refs.~\cite{Drees:2012fb,Christensen:2012ei,Christensen:2012si}.  
We will refer the above as ``{\bf Inclusive LEP-LHC Higgs}'' ({\bf ILLH}) 
scenario in this analysis.
After the Higgs discovery announcement in the summer of 2012, 
there have been a few studies on NMSSM\cite{NMSSMreviewetc} which could easily accommodate the 
above ILLH physics\cite{Belanger:2012tt,Cerdeno:2013cz}. 
There also have been a few analysis that reported results in 
multiple Higgs bosons near 125 GeV\cite{Gunion:2012gc} or a 
combined LHC plus Tevatron scenario of a 125~GeV plus a 136~GeV 
Higgs bosons\cite{Belanger:2012sd}, both in the context of NMSSM.
   
After the analyses of the Refs.\cite{Drees:2012fb,Christensen:2012ei,Christensen:2012si}, 
stronger constraints came from CMS\cite{CMSdataH2TauTau} and 
ATLAS\cite{ATLASdataH2TauTau} for the 
decay of neutral Higgs bosons $(H/A)$ to $\tau$ pairs.  
In MSSM, CMS data excludes parameter zones with 
$90 <M_A < 250$~GeV for $\tan\beta$ approximately above 5.5. As pointed out 
in Ref.\cite{Arbey:2012bp} there is a drastic reduction in the above 
limit of $\tan\beta$ for the same range of $M_A$ in comparison to 
the older result of 2011 \cite{cmsh2tautau2011}. This forces us to 
concentrate our 
search of the ILLH scenario for $\tan\beta \lsim 5.5$. 
On the other hand, we must take into account the 
ATLAS analyzed constraint coming from 
$t \bar t$ events, where $t \rightarrow b H^+$ with 
$H^+ \rightarrow \tau^+ \nu_\tau$\cite{Aad:2012tj}.  
This analysis indicates that, approximately for $\tan\beta$ between 2 to 6,   
parameter regions satisfying $90 < m_{H^+} < 150$~GeV become disallowed. 
Thus focusing on the two constraints 
from $A/H \rightarrow \tau^+ \tau^-$ 
and $H^+ \rightarrow \tau^+ \nu_\tau$, we look for 
the ILLH scenario for a small range of $\tan\beta$ namely 
$3<\tan\beta< 5.5$. The lower limit of $\tan\beta=3$ is chosen 
so as to be consistent with Ref.\cite{Aad:2012tj}\footnote{This 
is consistent with the LEP data analysis of 
Ref.\cite{Schael:2006cr} which however corresponds to an SM-like SUSY Higgs 
boson. For a 98 GeV non-SM like Higgs boson $h$, 
$\tan\beta$ can indeed be smaller than 3. Thus a much general 
analysis may probe smaller values of $\tan\beta$ than 3.}.
    With the range of $\tan\beta$ to be searched becoming restricted 
as mentioned above in the aforesaid non-decoupling region of MSSM for  
the ILLH scenario, one would require i) 
$\sin^2(\alpha-\beta)$ to be sufficiently small 
so as to have a small $Zh$ coupling 
in order to account for the 98~GeV data from LEP and ii) 
the heavier CP-even Higgs bosons $H$ 
to be SM-like and be consistent with the 
LHC signals in di-photon, 4-lepton and $W W^*$ channels. 
We also consider various low energy constraints. One of the 
most important constraints is 
$Br(B_s \rightarrow \mu^+ \mu^-)$\cite{bsmumuexperimental,bsmumurefs} 
simply because 
of its inverse quartic relationship with the pseudoscalar mass $M_A$. Besides, 
we explore the role of other relevant 
constraints like $Br(b \rightarrow s  \gamma)$\cite{Amhis:2012bh}  and 
$Br(B \rightarrow \tau \nu_\tau)$\cite{b2taunuSUSY,Lees:2012ju}. 
We will also impose the cold dark matter (CDM)\cite{dmreviews} 
constraint from PLANCK experiment\cite{Ade:2013lta}  
and compute the direct detection cross section 
of the lightest supersymmetric particle (LSP) for scattering with a proton 
in relation to the results of the XENON100 experiment\cite{Aprile:2012nq}. 
However, we ignore the constraint due to 
muon $g-2$ in this analysis because although the deviation from SM is large, 
there are uncertainties in the evaluations of the hadronic contributions 
leading to significantly varying final limits\cite{muong-2ETC}.
 
 We plan our analysis to proceed in the following way. In 
Section-\ref{MSSMdesc} we 
discuss the MSSM parameter scan strategy along with discussing 
different constraints to be 
imposed. In Section-\ref{alltheresults} we show the results of our scanning  
in MSSM parameter space 
in the light of the ILLH scenario along with a discussion
on the most significant constraints on the MSSM Higgs sector. In the
same section, we then move on to discuss the possibility of having 98 GeV 
as well as 125 GeV Higgs bosons in NMSSM. In Section-\ref{collidersection}, 
we discuss the collider implications of this scenario and finally in 
Section-\ref{conclusionsection} we summarize
our results.

\section{Probing MSSM parameter space for ILLH Scenario}
\label{MSSMdesc}
As mentioned earlier, although the MSSM Higgs sector is specified only by a 
few parameters like $M_A$ and $\tan\beta$, it is due to the large 
radiative corrections to the Higgs boson masses several MSSM parameters, 
particularly the ones associated with the stop sector become important. 
The radiative corrections to $\Delta m_h^2$ is quantified 
as follows\cite{Djouadi:2005gj,HiggsOrig1,HiggsOrig2}. 
\begin{equation}
\Delta m_h^2=\frac{3{\bar{m_t}}^4}{2\pi^2 v^2 \sin^2\beta}
\left[{\rm log}\frac{M^2_S}{{\bar{m_t}}^2} + \frac{X_t^2}{2 M_S^2}
\left(1-\frac{X_t^2}{6 M_S^2}\right) \right ].
\label{higgscor}
\end{equation}
Here,
$M_S=\sqrt{m_{\tilde t_1}  m_{\tilde t_2}}$,
$v=246$~GeV, $X_t=A_t-\mu \cot\beta$ and $\bar {m_t}$ refers to 
the running top-quark mass. $A_t$ stands for trilinear coupling for top-quark  
given at the electroweak scale. The running top-quark mass also includes 
QCD and electroweak corrections. One must however include the radiative 
corrections from sbottom and stau sectors for cases with 
large $\tan\beta$ and/or very large $\mu$. The importance of the latter  
comes from a ${|\mu|}^4$ dependence\cite{Altmannshofer:2012ks}. 
We will particularly scan MSSM parameter space 
reaching up to large values of $|\mu|$ without being 
concerned with any fine-tuning issue.  Hence, the contributions at least from 
the sbottom sector are hardly negligible. 
Expressions of radiative corrections for heavier neutral Higgs bosons and 
charged Higgs bosons may be seen in Ref.\cite{Djouadi:2005gj}. 

      With the above importance of Higgs related parameters 
of MSSM connected to radiative corrections for Higgs masses 
and the fact that we also would like to explore the parameter space that  
satisfy the XENON100 data\cite{Aprile:2012nq}, 
we select the following parameter ranges for 
scanning of MSSM parameter space. We point 
out that many of the parameters that are important to satisfy the 
CDM relic density\cite{dmreviews,AfewPapersOndm} 
constraints do not have significant effects on the 
Higgs sector or strongly interacting sectors that are relevant for 
Higgs\cite{Choudhury:2012tc}.  
For simplicity, we consider the squark masses of the first two 
generations as well as slepton masses of all the three generations 
sufficiently heavy. This will obviously not 
affect the Higgs masses whereas it will only avoid the possibility 
of LSP-slepton coannihilation scenarios for CDM constraint without 
any loss of generality.  We generate approximately 70 million random 
points in the following combined range of parameters.
\begin{eqnarray}
3<\tan\beta <5.5, ~0.085< M_A < 0.2 ~{\rm TeV}, ~0.3~{\rm TeV} <\mu < 
12~{\rm TeV},  \nonumber \\ 
0.05~{\rm TeV} < M_1,  M_2 < 1.5~{\rm TeV}, ~0.9~{\rm TeV} <M_3 <3~{\rm TeV}, \nonumber \\ 
-8 ~{\rm TeV} < A_t < 8 ~{\rm TeV}, ~-3 ~{\rm TeV} < A_b, ~A_\tau 
< 3 ~{\rm TeV}, ~A_u=A_d=A_e=0, \nonumber \\ 
0.3~{\rm TeV} <M_{\tilde q_3} <  5 ~{\rm TeV}, ~{\rm where}, ~{\tilde q_3} 
\equiv {\tilde t_L},{\tilde t_R},{\tilde b_L},{\tilde b_R} \nonumber \\ 
M_{{\tilde q}_i}= 3~{\rm TeV}, {\rm for}~i=1,2~~ {\rm and}~~ M_{{\tilde l}_i}=3~{\rm TeV}, {\rm for}~i=1,2,3.
\label{parameterRanges}
\end{eqnarray}
Among the SM parameters, we consider 
${m_b}^{\overline {\rm MS}}(m_b)=4.19$~GeV and $m_t^{\rm pole}=173.3 \pm 2.8$~GeV. 
$m_t^{\rm pole}$ is varied within the above range following the argument of Ref.\cite{Alekhin:2012py}. 
We scan the parameters 
within the above ranges of Eq.\ref{parameterRanges} 
while imposing lower limits on the sparticle masses\cite{Beringer:1900zz} 
and eliminating the possibility of having any charge and color breaking (CCB) 
minima\cite{casasCCB}. 

   While sketching out the valid parameter space we also consider the 
errors in the masses of two CP-even neutral Higgs bosons. 
We consider a theoretical uncertainty amount of 3~GeV in the Higgs mass 
computation that arises out of uncertainties in the renormalization scheme, 
scale dependence, the same in higher order loop corrections up to three 
loops or that due to the top-quark 
mass\cite{Arbey:2012dq,Heinemeyer:2011aa,Allanach:2004rh,Degrassi:2002fi,higgs3loop}. 
Thus, we isolate the parameter space 
with the following limits for the Higgs boson masses.
\begin{eqnarray}
95~{\rm GeV} < m_h < 101~{\rm GeV}, ~{\rm and}\nonumber \\
122~{\rm GeV} < m_H < 128~{\rm GeV}.
\label{lep-lhc-masslimits}
\end{eqnarray}
We further require $h$ to have 
non-SM like couplings by demanding $\sin^2(\beta-\alpha)$ to be small. 
The observed LEP excess\cite{Barate:2003sz} approximately results into: 
\begin{equation}
0.1<\sin^2(\beta-\alpha)<0.25.
\label{LEPcriterion}
\end{equation} 
\noindent
We now focus on the primary production channels of Higgs boson at the LHC 
and define the following ratios related to the gluon-gluon fusion 
and the vector boson fusion channels.
\begin{equation}
R^{h,H}_{gg}(XX) = \frac{\Gamma(h,H \rightarrow g g) {\rm Br}(h,H \rightarrow XX)}
{\Gamma(h_{SM} \rightarrow g g) {\rm Br}(h_{SM} \rightarrow XX)}
\end{equation}
and 
\begin{equation}
R^{h,H}_{Vh/H}(YY) = \frac{\Gamma(h,H \rightarrow WW) {\rm Br}(h,H \rightarrow YY)}
{\Gamma(h_{SM} \rightarrow W W) {\rm Br}(h_{SM} \rightarrow YY)}.
\end{equation}
where, $XX = \gamma\gamma$ or $ZZ^*/ W^+W^-$ and  
$YY = b\bar b$ or $\tau^+ \tau^-$. 
As we know that the production of Higgs at the LHC is dominated 
by gluon-gluon fusion both in the context of SM, 
as well as in our parameter zone of interest of MSSM.   
However, detection of Higgs bosons via 
$h,H\rightarrow  b \bar{b}$ and $h,H\to\tau^+\tau^-$ 
decays becomes 
nearly impossible when the Higgs boson is produced via gluon-gluon
fusion since it is likely to be overshadowed by di-jet events from 
QCD interactions. On the other hand, when Higgs boson is 
produced in association with a 
vector boson ($W/Z$), also known as Higgs-strahlung (HS), with 
the gauge bosons decaying 
leptonically, it is relatively easy to tag the two $b$-jets 
to reconstruct the Higgs mass. Both ATLAS and CMS have probed 
Higgs signatures 
in this channel but with large uncertainties, which one would expect to 
be modified with enough data collected at the 14~TeV LHC run.  
Hence, we consider the Higgs-strahlung process when computing 
$R^{h,H}_{Vh/H}(b \bar b)$ for Higgs boson signals 
decaying into pair of bottom quark.  
Because of large uncertainties in these channels
we refrain ourselves from imposing any limit on 
$R_{VH}^H(b \bar b)$ or $R_{VH}^H(\tau^+ \tau^-)$ in our analysis. 
Rather, 
we only set a lower limit for $R_{gg}^H(\gamma \gamma)$: 
\begin{equation}
R_{gg}^H(\gamma \gamma)>0.5.
\label{gammagammaexcess}
\footnote{The most significant change in the recent updates 
on LHC results at the Moriond-2013 conference
is that CMS no longer observes the excess in the di-photon channel
from their multivariate analysis. However, a small excess is still
there in their cut-based analysis. Moreover, ATLAS observes some excess 
in the di-photon mode (see Table~\ref{HiggsDecayTableData} for details). 
Therefore, we choose a conservative limit of $R_{gg}^H(\gamma \gamma)>0.5$
which is within 1$\sigma$ of the CMS di-photon result (multivariate analysis).}
\end{equation}
    We now consider various low energy constraints. The experimental 
data from $Br(b \rightarrow s \gamma)$ almost saturates the SM value that 
comes from $t-W$ loop\cite{bsgammaSMoriginals}. Within MSSM, the dominant 
contributions  
come from $t-H^\pm$ and ${\tilde t}_{1,2}-{\tilde \chi}_{1,2}^\pm$ 
loops\cite{bsgammaSUSYorigEtc}, the former having the same 
sign as that of the $t-W$ loop of SM. The chargino loop contribution is 
proportional to $A_t \mu \tan\beta$.  Depending on the sign of $A_t \mu$, 
there can be  
cancellation or enhancement between the above 
loop contributions within MSSM\cite{Carena:2000uj}. 
Furthermore, we note that in SUSY,   
$Br(b \rightarrow s  \gamma)$ may have an indirect but important   
contribution from a gluino-squark loop at the next-to-leading order 
level. 
This is primarily connected to the renormalization of Yukawa couplings to 
down type of fermions. The corrections that can be summed to all order  
in perturbation theory 
relates to the 
supersymmetric quantum chromodynamic (SQCD) 
corrections to the mass of the bottom quark $m_b$ and this is proportional 
to $\mu M_{\tilde g}\tan\beta$\cite{Carena:2000uj,Carena:1999py}. 
This leads to 
alteration of the $\bar t b H^+$ vertex leading to a correction 
to $Br(b \rightarrow s \gamma)$
that comes with an opposite 
sign with respect to the leading order contribution of the  
$t-H^\pm$ loop\cite{Carena:2000uj}. 
Typically, the correction is seen as a required one for 
large values of $\tanbeta$. But in spite of $\tan\beta$ being 
small in our analysis, the same correction is also 
very important because of large possible values for $\mu$\cite{Carena:1999py}.
This next-to-leading order effect can 
potentially cancel the leading order effects from 
chargino or charged Higgs loops\footnote{SUSY 
electroweak corrections to bottom Yukawa couplings may also 
be somewhat appreciable for very large values of $\mu M_2$\cite{Carena:1999py}.}. 
The experimental 
result is $Br(b \rightarrow s \gamma)=(343 \pm 22)\times 10^{-6}$\cite{Amhis:2012bh} 
which at 3$\sigma$ level leads to: 
\begin{equation}
2.77 \times 10^{-4}  <Br(b \rightarrow s  \gamma)<4.09 \times 10^{-4}.
\label{bsgammalimits}
\end{equation}
    Next, as mentioned before, we must explore the effect of the limits from 
$B_s \rightarrow \mu^+ \mu^-$ particularly because of its $M_A^{-4}$ 
dependence while we probe a light $M_A$ scenario. The 
experimental (LHCb) and SM values are given by 
$Br{(B_s \rightarrow \mu^+ \mu^-)}_{exp}=(3.2^{+1.4}_{-1.2}(\mbox{stat.}) ~^{+0.5}_{-0.3}   
(\mbox{syst.})) \times 10^{-9}$\cite{bsmumuexperimental}. 
This is to be compared with the SM 
result $Br(B_s \rightarrow \mu^+ \mu^-)_{SM}=(3.23 \pm 0.27) 
\times 10^{-9}$~\cite{Buras:2012ru}. As in Ref.\cite{Roszkowski:2012nq}, 
we combine the errors of LHCb data and SM result 
to obtain the following $2\sigma$ limits.
\begin{equation}
0.67 \times 10^{-9} <Br(B_s \rightarrow \mu^+ \mu^-)<6.22 \times 10^{-9}.
\label{bsmumulimit}
\end{equation}
We also take into account the combined constraint from 
$Br(B^+ \rightarrow \tau^+ \nu_\tau)$ from BABAR\cite{Lees:2012ju} 
and $Br(B^- \rightarrow \tau^- \bar{\nu}_\tau)$ from 
Belle\cite{Adachi:2012mm}. With 
$R_{(B \rightarrow \tau \nu_\tau)}=
\frac{Br{(B \rightarrow \tau \nu_\tau)}_{SUSY}}{Br{(B \rightarrow \tau
\nu_\tau)}_{SM}}$, we find 
$0.31 < R_{(B \rightarrow \tau \nu_\tau)}<2.10$ 
(see Ref.\cite{Chakraborti:2012up} for details). Typically, the above 
constraint may be effective for large $\tan\beta$ and smaller charged Higgs 
boson mass. 
The limits of $R_{(B \rightarrow \tau \nu_\tau)}$ however do not impose any 
significant constraint on the parameter space of our study. 

\noindent
Finally, we take into account the CDM relic density limits recently 
provided by the  
PLANCK\cite{Ade:2013lta} collaboration, which 
at $3\sigma$ level reads,
\begin{equation}
0.112  < \Omega_{{\widetilde \chi}_1^0}h^2<0.128.
\label{planckdata}
\end{equation}
We also compute spin-independent 
direct detection LSP-proton scattering cross-section and compare 
with the XENON100 data\cite{Aprile:2012nq}. The spin-independent 
cross section depends on t-channel Higgs and s-channel squark 
exchanges of which the former dominates unless the squark masses 
are close to that of the LSP\cite{Drees:1993bu,dmmany}. We note that in a thermal 
production scenario of dark matter we may have the following possibilities: 
i) either the lightest neutralino constitutes the entire 
CDM relic density while corresponding to 
correct relic abundance satisfying PLANCK data or ii) it is just one of 
the candidates of a multicomponent DM combination with obviously 
a smaller relic density contribution from its own, 
below the lower limit of Eq.\ref{planckdata}. For the 
latter scenario one has $\Omega_{{\widetilde \chi}_1^0} h^2 < {(\Omega_{CDM} h^2)}_{\rm min}$, 
where ${(\Omega_{CDM} h^2)}_{\rm min}$ refers to the lower limit 
of Eq.\ref{planckdata}. Here, we should multiply 
with the fraction $\zeta=\rho_\chi/\rho_0$ the LSP 
contributes to the total local density of DM to obtain the true event rate 
where $\rho_0$ denotes the 
local total DM density and $\rho_\chi$ means the DM density contributed 
by the LSP. 
Thus, one conveniently defines the ratio as  
$\zeta={\rm min}\{1,\Omega_{{\widetilde \chi}_1^0} h^2/{(\Omega_{CDM} h^2)}_
{\rm min}\}$\cite{BottinoRescaled}. $\zeta$ by definition is 
unity for scenarios with 
right abundance (single component) or over-abundance of DM.

\section{Results}
\label{alltheresults}

\subsection{MSSM}
\label{mssmsubsectionforresults}
We search for the ILLH scenario 
within the phenomenological MSSM parameter space 
corresponding to Eq.\ref{parameterRanges}.
We use the code SuSpect (version 2.41)\cite{SuspectCode} for 
generating sparticle spectra and 
micrOMEGAs (version 2.4.5)
\cite{Belanger:2008sj,Belanger:2005kh,Belanger:2006is} to 
compute neutralino relic density, LSP-nucleon 
direct detection cross-section as well as various B-physics 
related quantities. 
We compute relevant Higgs decays by using the code 
SUSY-HIT\cite{Djouadi:2006bz}. 

\begin{figure}[!htb]
\vspace*{0.3in}
\begin{center}
\includegraphics[scale=0.45]{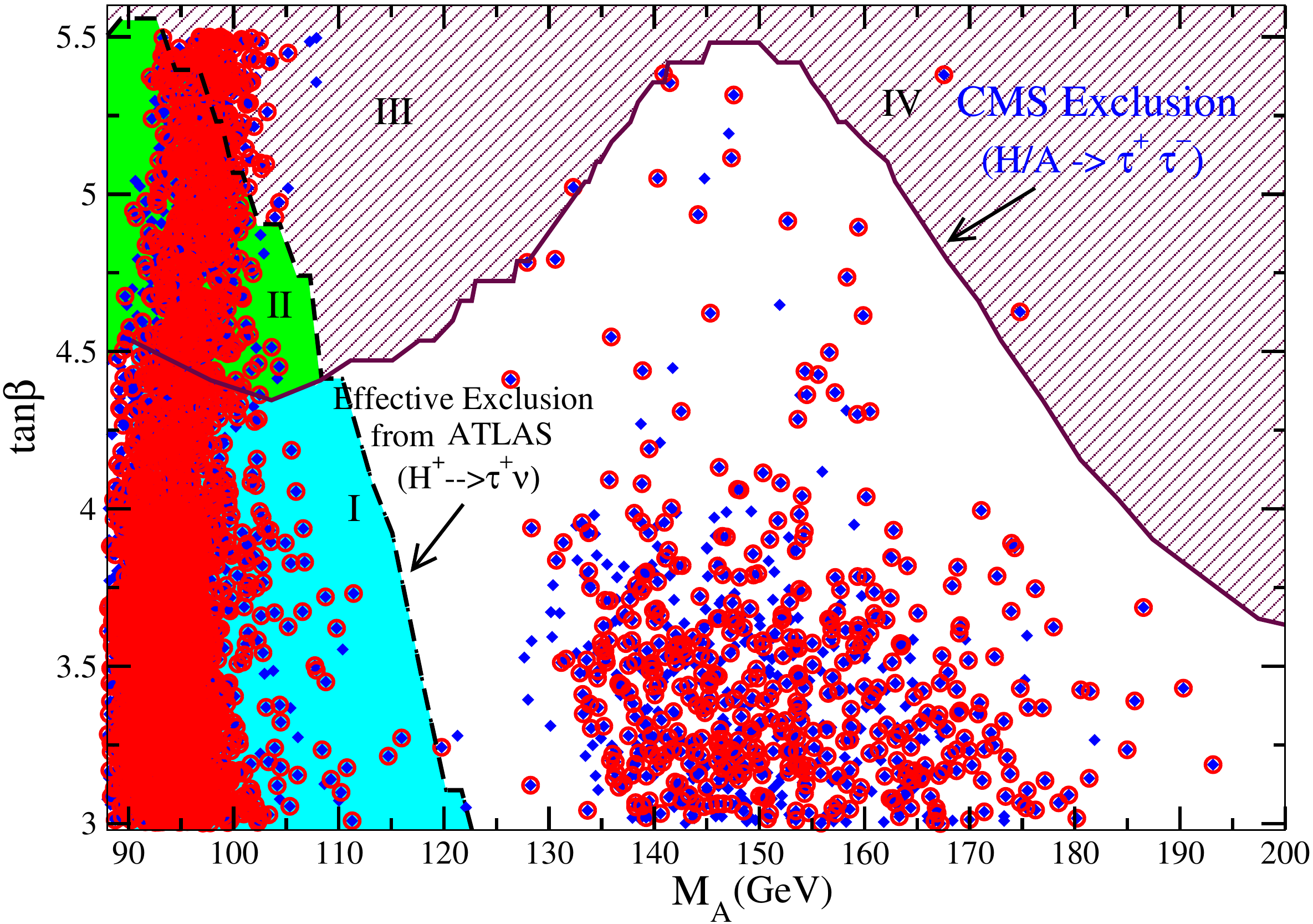}
\caption{{\small 
{\it
Scatter plot of mass of pseudoscalar Higgs boson $M_A$ vs $\tan\beta$ in 
the ILLH scenario of MSSM.
The diamond (blue) shaped points satisfy the constraints of 
Eqs.\ref{lep-lhc-masslimits} to \ref{bsmumulimit}. The 
(red) circles (enclosing diamonds) additionally satisfy the DM relic 
density constraint. The shaded region above the 
solid (maroon) line is discarded via 
CMS data of $H/A \rightarrow \tau^+ \tau^-$.  
The region below the dashed line is discarded 
via the indirect effect 
of the $H^+ \rightarrow \tau^+ \nu_\tau$ constraint 
from ATLAS. This is done by translating the constraint 
into the $M_A-\tan\beta$ plane by using the 
tree level relationship between the masses of pseudoscalar and 
charged Higgs bosons in MSSM. 
}
}
}
\label{MA-tanbeta}
\end{center}
\end{figure}

\begin{figure}[!htb]
\begin{center}
   \includegraphics[scale=0.2]{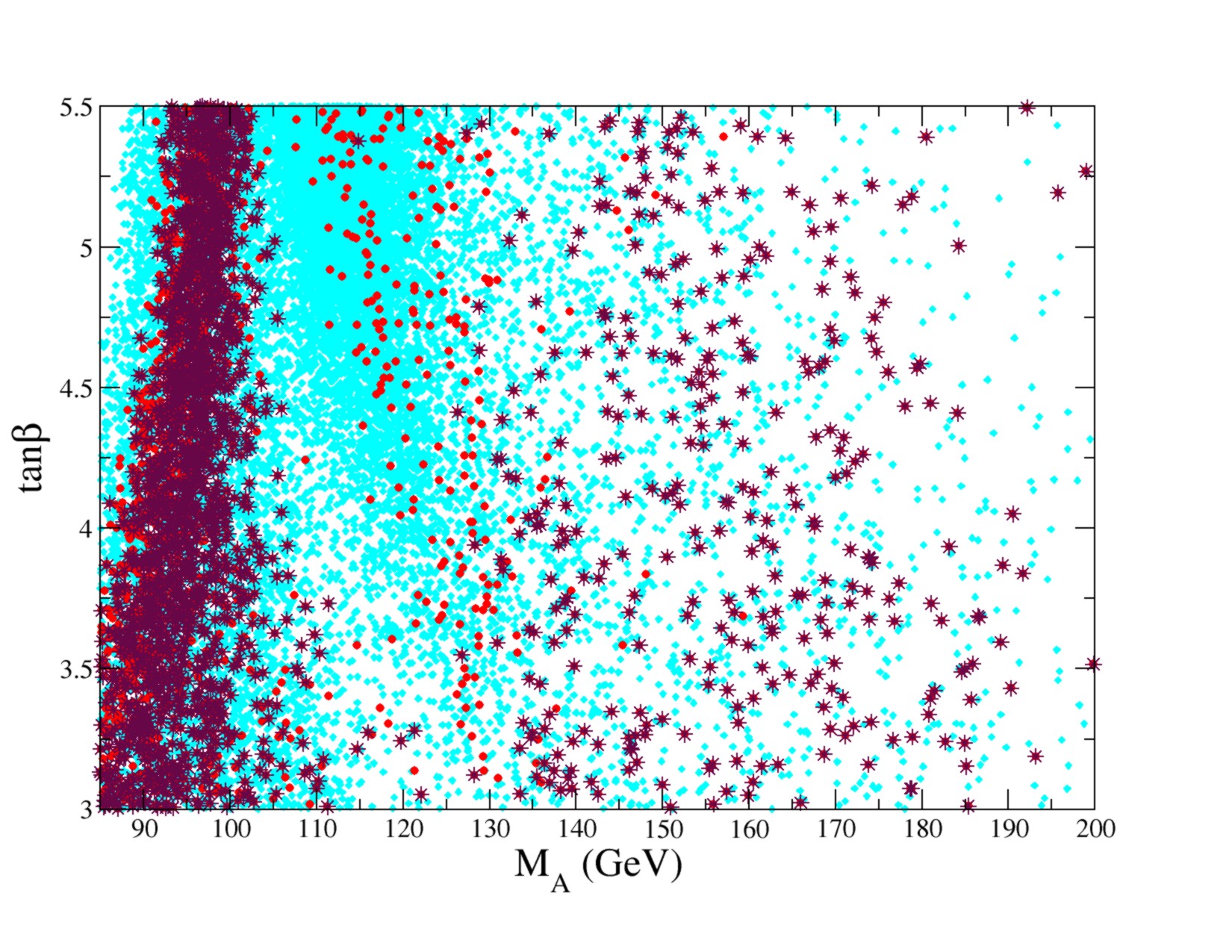}
\vspace*{-0.8cm}
\caption{
{\small
{\it Scatter plot of mass of pseudoscalar Higgs boson $M_A$ vs $\tan\beta$ in 
the ILLH scenario of MSSM.  
The lightest (cyan) scattered points correspond to parameter points 
that at least  
satisfy the two Higgs mass constraints of Eq.\ref{lep-lhc-masslimits}.
Dark (red) filled circles correspond to imposing 
$\sin^2(\beta-\alpha)$ limits from Eq.\ref{LEPcriterion}. 
The star (maroon) shaped points are obtained by imposing 
$Br(B_s \rightarrow \mu^+ \mu^-)$ limits from Eq.\ref{bsmumulimit}. 
}}
}
\label{MA-tanbeta-Constraints}
\end{center}
\end{figure}

In Fig.\ref{MA-tanbeta} we show 
the result of our parameter space scanning in $M_A-\tan\beta$ plane. 
The results 
show parameter points in the above plane that satisfy 
i) the sparticle mass lower limits including those from 
LHC, ii) the 
two neutral CP-even Higgs boson masses $m_h$ and $m_H$ to be within the 
ranges 
of Eq.\ref{lep-lhc-masslimits}, iii) the LEP specified limits for $h$-boson  
to be non-Standard Model like as in Eq.\ref{LEPcriterion} and iv) 
the chosen value $R_{gg}^H(\gamma \gamma)>0.5$. 
Besides the above, 
the parameter points also satisfy limits for $Br(b \rightarrow s  \gamma)$ as 
shown in Eq.\ref{bsgammalimits} and the LHCb limits of 
$Br(B_s \rightarrow \mu^+ \mu^-)$ of Eq.\ref{bsmumulimit}. 
We have also verified the validity of the range of 
$R_{(B \rightarrow \tau \nu_\tau)}$ as mentioned 
in the last section. The scattered points 
that satisfy the above limits are shown with diamond label (in blue). 
The red circles (enclosing diamonds) additionally 
show the points that satisfy 
the DM relic density constraint. In this analysis we consider 
the possibility of LSP to be also a 
sub-dominant candidate of dark matter. Thus consistency with the CDM
relic density constraint here means satisfying only the 
upper limit of Eq.\ref{planckdata}. 
There is a general absence of valid points in the region of 
$110 \lsim M_A \lsim 125$~GeV. We will come back to this point in 
the discussion of Fig.\ref{MA-tanbeta-Constraints}. 

The recent results of 
decay of heavy neutral Higgs bosons $H/A$ 
to $\tau$ pairs from CMS\cite{CMSdataH2TauTau} 
and ATLAS\cite{ATLASdataH2TauTau} keep only the small $\tan\beta$ zone 
to be allowed for small $M_A$ region ($< 200$~GeV). 
Because of the above data there is no allowed zone for 
$\tan\beta> 5.5$ for $M_A<200$~GeV as shown by the black solid line 
of Fig.\ref{MA-tanbeta}. 
The dashed line, on the other hand, shows the indirect effect 
of $H^+ \rightarrow \tau^+ \nu_\tau$\cite{Aad:2012tj} constraint 
from ATLAS when translated into the $M_A-\tan\beta$ plane by using the 
tree level relationship between the masses of pseudoscalar and 
charged Higgs bosons in MSSM, namely $M_{H^\pm}^2=M_A^2+M_W^2$.  
The latter constraint eliminates $M_A$ below 
120~GeV. The regions marked by I,II,III and IV are thus discarded via 
$H/A \rightarrow \tau^+ \tau^-$ and $H^+ \rightarrow \tau^+ \nu_\tau$ 
data. Hence, the surviving $M_A$ range for $3< \tan\beta <5.5$, 
as shown in Fig.\ref{MA-tanbeta} is 
given by,
\begin{equation}
130<M_A<200 \rmGeV.
\label{survivingMA} 
\end{equation}
\vspace*{-1.1cm}

We now discuss the details of the interplay of the most important 
constraints of our analysis in the 
ILLH scenario of MSSM in the $M_A-\tan\beta$ plane in 
regard 
to the blanck intermediate zone 
$110 \lsim M_A \lsim 125$~GeV as mentioned above.
The lightest (cyan) scattered points of 
Fig.\ref{MA-tanbeta-Constraints} correspond to parameter points that at least 
satisfy the two Higgs mass constraints of Eq.\ref{lep-lhc-masslimits}.
The points in general span the $M_A-\tan\beta$ plane, albeit with 
varying degree of existence for different zones. Imposing the LEP Higgs information 
regarding $\sin^2(\beta-\alpha)$ from Eq.\ref{LEPcriterion} (shown as 
darker (red) circles) drastically 
reduces the available parameter space in the aforesaid intermediate 
zone of $M_A$. Further reduction of parameter points in the same region 
occurs via the constraint of $Br(B_s \rightarrow \mu^+ \mu^-)$ from 
Eq.\ref{bsmumulimit}. The star (maroon) shaped points refer to parameter 
points in $M_A-\tan\beta$ plane that 
satisfy Eq.\ref{bsmumulimit} in addition to Eqs.\ref{lep-lhc-masslimits} and 
\ref{LEPcriterion}. 
Thus the above step-by-step imposition of various constraints shows 
that requirement of $h$-boson to be non-SM like via 
Eq.\ref{LEPcriterion} and the stringent limits from 
$Br(B_s \rightarrow \mu^+ \mu^-)$  
(Eq.\ref{bsmumulimit}) in combination cause  
the absence of valid points in the intermediate $M_A$ 
zone (Fig.\ref{MA-tanbeta}).  
\begin{figure}[!htb]
\vspace*{0.3in}
\begin{center}
\includegraphics[scale=0.5]{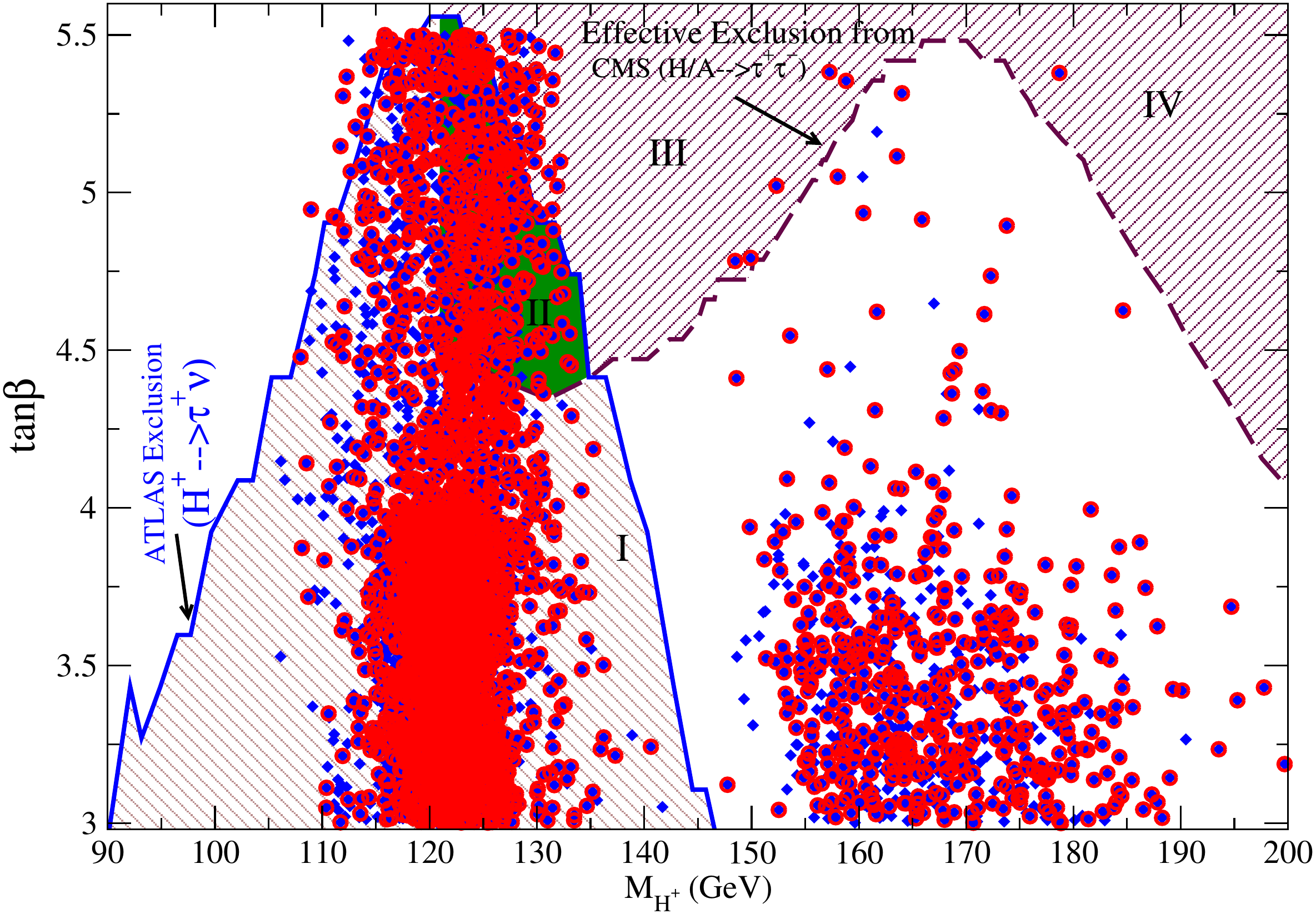}
\caption{{\small
{\it Scatter plot of mass of Charged Higgs boson 
$M_{{H}^\pm}$ vs $\tan\beta$ in the ILLH scenario of MSSM.
Symbols have the same meaning as in Fig.\ref{MA-tanbeta}.
Direct constraint from $H^+ \rightarrow \tau^+ \nu_\tau$ from 
ATLAS is shown as a blue solid line. 
Indirect effect of $H/A \rightarrow \tau^+ \tau^-$ from CMS is additionally 
drawn as a dashed (maroon) line by using the tree level 
relationship between $M_A$ and $M_{H^\pm}$. All the shaded regions are 
discarded via the above two constraints.   
}
}
}
\label{Mch-tanbeta}
\end{center}
\end{figure}\\
\noindent
    Fig.\ref{Mch-tanbeta} shows a scatter plot in $M_{H}^\pm-\tan\beta$ 
plane from same analysis. With 
similarly marked scattered points for the constraints as described above in 
connection with Fig.\ref{MA-tanbeta}, we draw direct constraint from 
$H^+ \rightarrow \tau^+ \nu_\tau$ from ATLAS\cite{Aad:2012tj} as a blue solid line. The 
indirect effect of $H/A \rightarrow \tau^+ \tau^-$ as seen in Fig.\ref{MA-tanbeta} is additionally 
drawn as a dashed (maroon) line by using the tree level 
relationship between $M_A$ and $M_{H^\pm}$. The region of $M_H^\pm <145$~GeV 
becomes entirely disallowed via $H^+ \rightarrow \tau^+ \nu_\tau$ from ATLAS. 
The combined disallowed zones are shown as regions I to IV. 
The valid 
zone of charged Higgs mass is seen to be
 $150~{\rm GeV} <M_{H^\pm}<200$~GeV that 
falls below the line corresponding to $H/A \rightarrow \tau^+ \tau^-$ 
data from CMS\cite{CMSdataH2TauTau}.

\subsubsection{Interplay of parameters related to Higgs mass radiative 
corrections: identifying most significant constraints}
We now consider studying the parameters connected to the Higgs mass 
radiative correction, that becomes highly relevant for the ILLH scenario. As discussed before, important corrections come from the top-stop as well as 
bottom-sbottom loops. The latter loops become important because of  
large range of values of $\mu$ considered in this analysis (Eq.\ref{parameterRanges}). We will also see that effect of considering 
large $\mu$ has an important consequence on the 
constraints of $Br(b \rightarrow s  \gamma)$. 
\begin{figure}[!htb]
\vspace*{-0.2in}
\mygraphh{MA-At-forConstraints}{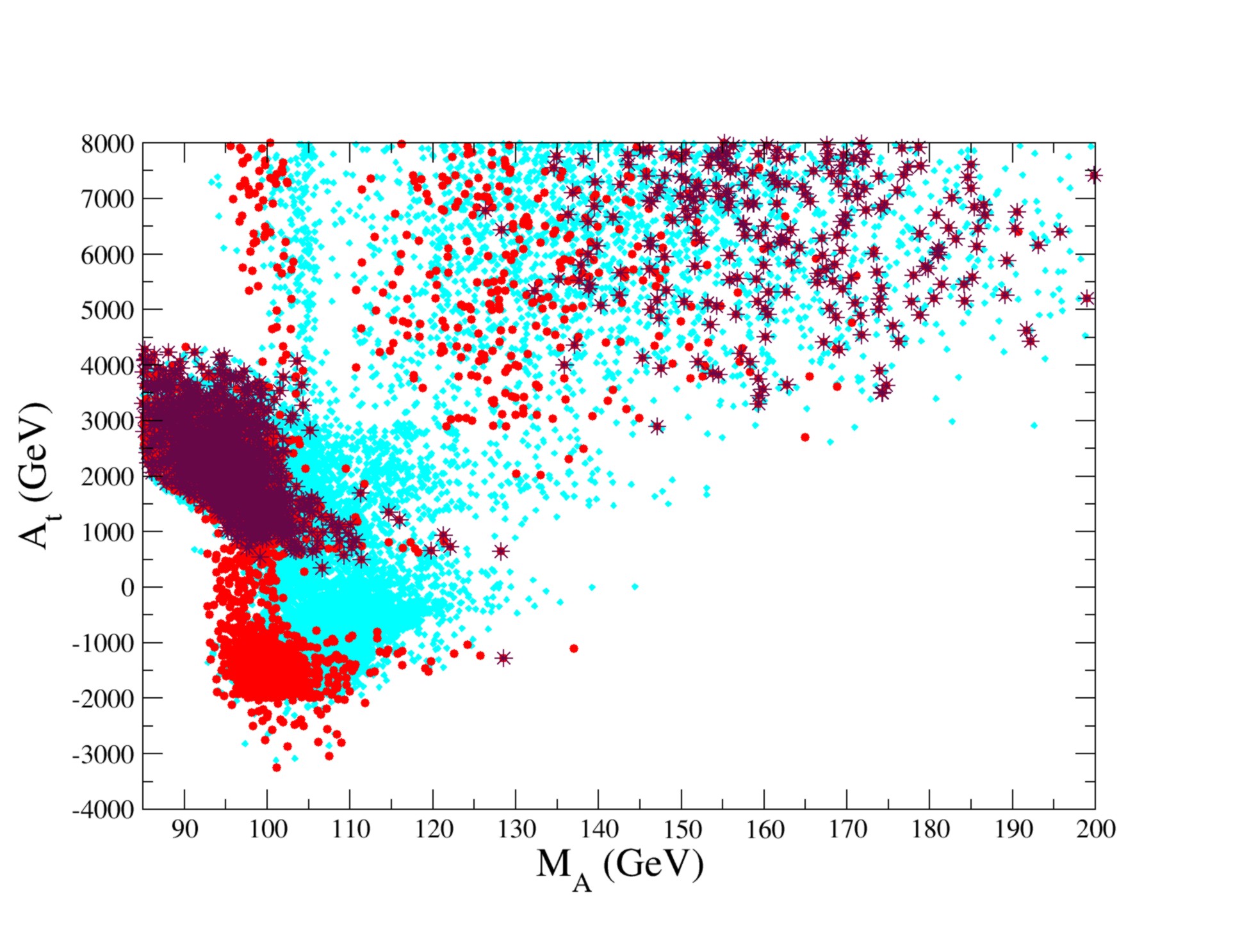}
\hspace*{0.5in}
\mygraphh{mA-mu-forConstraints}{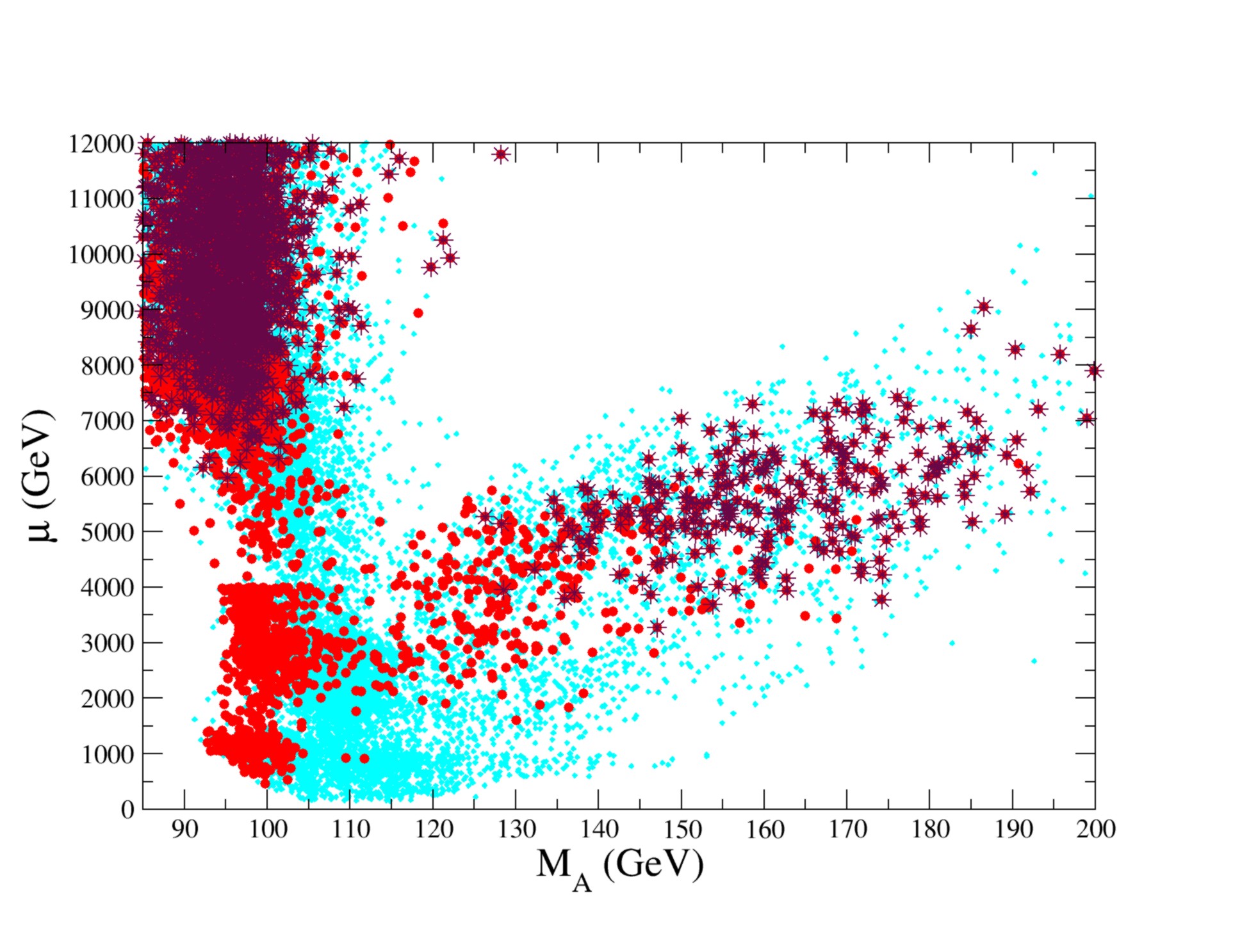}
\caption
{\small
{\it
a) Scatter plot of mass of pseudoscalar Higgs boson $M_A$ vs 
$A_t$ in the ILLH scenario of MSSM. 
All the lightest (cyan) points satisfy  
at least the Higgs mass limits of Eq.\ref{lep-lhc-masslimits} and allow 
$A_t$ to have either signs. Darker (red) circles additionally satisfy 
$\sin^2(\beta-\alpha)$ limits from Eq.\ref{LEPcriterion}. 
 Star (maroon) marked points refer to further imposition 
of limits from $Br(b \rightarrow s  \gamma)$  and 
$Br(B_s \rightarrow \mu^+ \mu^-)$ corresponding to 
Eqs.\ref{bsgammalimits} and \ref{bsmumulimit} respectively. This  
effectively discards negative values of $A_t$. 
b) Scatter plot of $M_A$ vs $\mu$ similarly obtained as 
Fig.\ref{MA-At-forConstraints} with having same meaning for the symbols.
}}
\label{At-mu-forConstraints}
\end{figure}\\
Fig.\ref{MA-At-forConstraints} shows scattered points in the $M_A-A_t$ plane. 
All the lightest (cyan) points satisfy  
at least the Higgs mass limits of Eq.\ref{lep-lhc-masslimits}. These points 
include both positive as well as negative values of $A_t$. 
We see that 
there are parameter points satisfying just the two Higgs mass limits 
for negative values of $A_t$ when $M_A$ spans  
approximately up to 130 GeV. On the other hand, only positive solutions of 
$A_t$ are available that satisfy 
the two constraints for $M_h$ and $M_H$ for 
larger values of $M_A$ that are consistent with 
Eq.\ref{survivingMA}.    
Further imposition of the LEP Higgs information 
regarding $\sin^2(\beta-\alpha)$ from Eq.\ref{LEPcriterion} results into 
darker (red) circles which are also understood to coexist with 
star (maroon) marked points that additionally obey 
the constraints of $Br(b \rightarrow s  \gamma)$  and 
$Br(B_s \rightarrow \mu^+ \mu^-)$ 
(Eqs.\ref{bsgammalimits} and \ref{bsmumulimit}).   
Fig.\ref{mA-mu-forConstraints} shows the effect in 
$M_A-\mu$ plane while constraints are imposed step-by-step as in 
Fig.\ref{MA-At-forConstraints} with the symbols having the same meaning.
Clearly, satisfying the two Higgs limits is possible both for small and large 
values of $\mu$. However like Fig.\ref{MA-At-forConstraints} 
constraints from  $Br(b \rightarrow s  \gamma)$  and $Br(B_s \rightarrow \mu^+ \mu^-)$ 
discard a large region of parameter space.
For the surviving zone of Eq.\ref{survivingMA} 
the above selects a 
zone with relatively small values of $\mu$ (up to 6~TeV ) 
(Fig.\ref{mA-mu-forConstraints}) as valid parameter range. 
Investigation reveals that it is indeed the limits from 
$Br(b \rightarrow s  \gamma)$ 
that disallow a large region of parameter space and we try to explain 
this in a brief {\it qualitative} detail by dividing the span of $M_A$ into 
two distinct zones {\bf (i) below} and {\bf (ii) above} 130 GeV for both the subfigures 
of Fig.\ref{At-mu-forConstraints}. 
The above is same as dividing into smaller and 
relatively larger $M_{H^\pm}$ zones. We remind that 
$t-H^\pm$ loop contributes to $Br(b \rightarrow s  \gamma)$  
with the same sign as that of $t-W$ loop of SM\cite{Carena:2000uj}. \\
\bul{\bf Smaller $M_A$ zone ($\lsim 130$~GeV) with $A_t<0$ where 
$Br(b \rightarrow s  \gamma)$ exceeds the allowed limit:}\\
Focusing on the smaller half side of  
$M_A$ (i.e. below 130 GeV) of Fig.\ref{MA-At-forConstraints}, analysis 
revealed that 
$Br(b \rightarrow s  \gamma)$ typically 
becomes larger than the upper limit of 
Eq.\ref{bsgammalimits} when $A_t<0$ (but there 
are some cases where the value goes below the lower bound too in our 
multi-dimensional parameter space of MSSM, as we will see below). 
It turns out that the same smaller 
$M_A$ region of Fig.\ref {mA-mu-forConstraints} ({i.e.} the lower 
half side) that is also associated with negative values of $A_t$ and 
for which $Br(b \rightarrow s  \gamma)$ exceeds the 
allowed limit 
typically corresponds to smaller values of $\mu$. 
Thus combining the results of the two 
subfigures of Fig.\ref{At-mu-forConstraints} we find that for most of the 
parameter zones corresponding to the lower half of $M_A$ and negative 
$A_t$, the values of $|\mu A_t|$ are relatively small since both 
the components namely  $\mu$ and $|A_t|$ 
are individually on the smaller side. Thus in spite of having $\mu A_t<0$, an  
apparently favorable situation for cancellation of diagrams in 
$Br(b \rightarrow s  \gamma)$ analysis, 
the chargino-stop loop contribution is not sufficient to overcome the 
large charged Higgs contribution which is typically larger in the smaller  
$M_A$ zone than what it would be for the larger $M_A$ zone as described 
above. Thus $Br(b \rightarrow s  \gamma)$ limit is violated in its 
upper bound. \\
\bul{\bf Smaller $M_A$ zone with $A_t<0$ where 
$Br(b \rightarrow s  \gamma)$ is below the lower limit:}\\
As mentioned before, in the low $M_A$ zone, additionally 
there are small regions of parameter 
space with negative $A_t$ where 
$Br(b \rightarrow s  \gamma)$ may become 
lesser than the lower limit of Eq.\ref{bsgammalimits} in the region of 
a sufficiently large $\mu$.  
Thus chargino contribution may significantly neutralize the charged Higgs 
contribution but as mentioned in Section-\ref{MSSMdesc} there 
may be an important gluino-squark loop contribution 
which comes with an opposite sign wrt the charged Higgs contribution. The 
latter contribution which goes in tandem with the negative contribution 
from the chargino loop (while $A_t<0$) causes  
$Br(b \rightarrow s  \gamma)$ to go 
below the lower limit of Eq.\ref{bsgammalimits}.  \\
\bul{\bf Smaller $M_A$ zone with $A_t>0$:}\\
Here the chargino and the charged Higgs loop contributions add up 
but the gluino contribution which is important for large $\mu$ and/or large 
$M_{\tilde g}$ has an opposite sign. 
Even with varying stop, chargino and gluino 
masses, parameter points with very large and positive values of $A_t$  
typically have exceedingly large values of $Br(b \rightarrow s  \gamma)$ 
unless $\mu$ becomes sufficiently large to cause a large degree 
of cancellation via the gluino loop. Such cancellations are indeed indicated 
in both the panels 
of Fig.\ref{At-mu-forConstraints} for the low $M_A$ zone when the 
cluster of star marked points span from  $0 <A_t< 4$~TeV whereas $\mu$ 
varies from 6 to 12 TeV satisfying both the limits 
of $Br(b \rightarrow s  \gamma)$ of Eq.\ref{bsgammalimits}.\\
\bul{\bf Larger $M_A$ zone ($\gsim 130$~GeV) where $A_t$ is essentially 
all positive:}
Here the parameter points that satisfy the two Higgs mass limits 
are such that $A_t$ is essentially always positive (except 
an insignificant number of parameter points). The charged Higgs mass 
goes higher, 
hence its contribution is relatively small. Although $A_t$ is in the 
larger side, $\mu$ is relatively in the moderate zone. Thus the 
combined chargino plus charged Higgs contribution which have the same signs 
is not exceedingly large and is balanced by the gluino contribution which 
comes with a negative sign. Thus in this zone of $M_A$,  
$Br(b \rightarrow s  \gamma)$ constraint is likely to be satisfied a fact 
which is consistent with the figures. It is also quite likely to 
have adjacent parameter points that would fall below or above the 
limits of $Br(b \rightarrow s  \gamma)$ of Eq.\ref{bsgammalimits} in this 
zone of $M_A$. 
\begin{figure}[!htb]
\vspace*{0.3in}
\mygraph{ma_mu_actual}{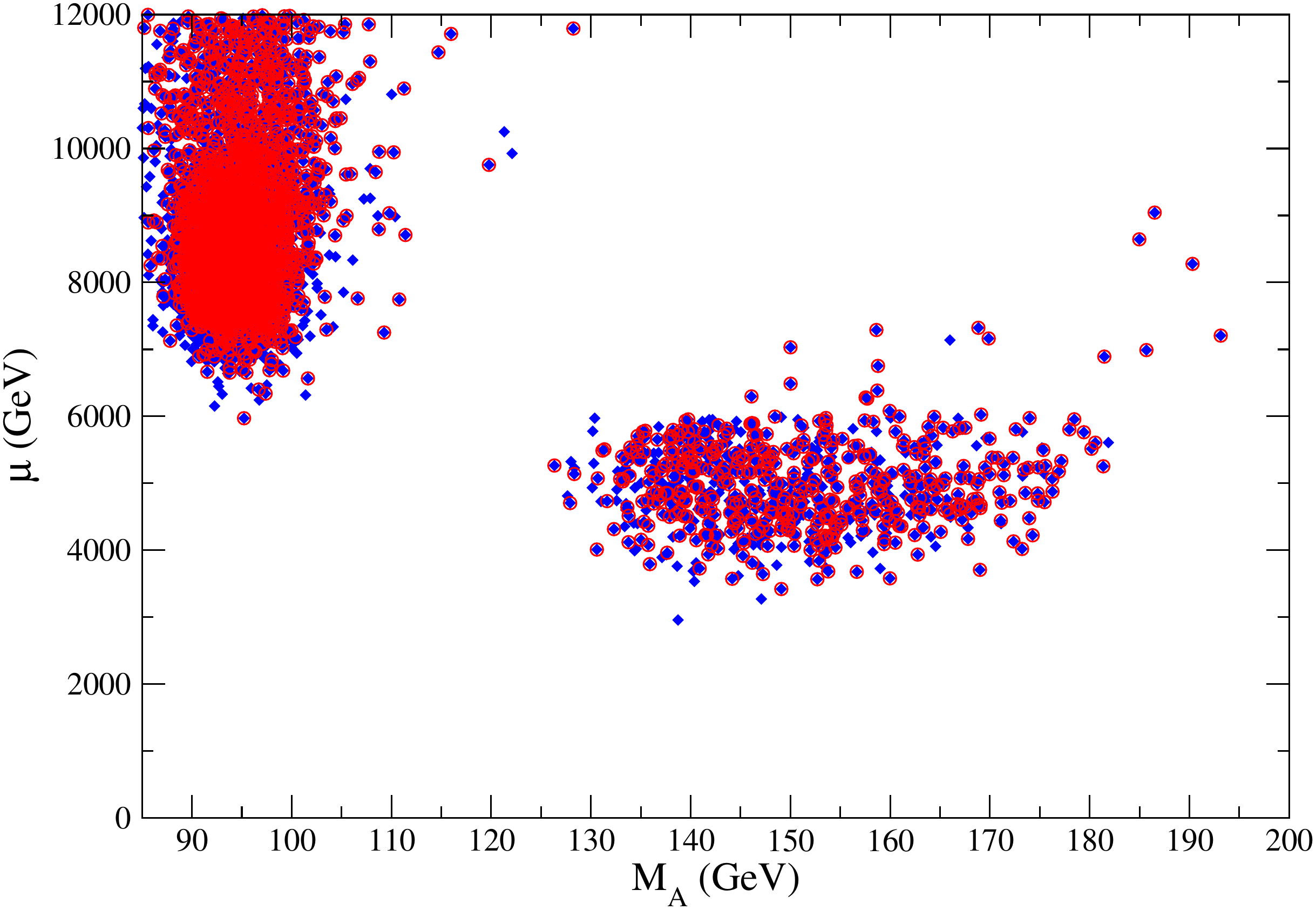}
\hspace*{0.5in}
\mygraph{ma_vs_mstop_actual}{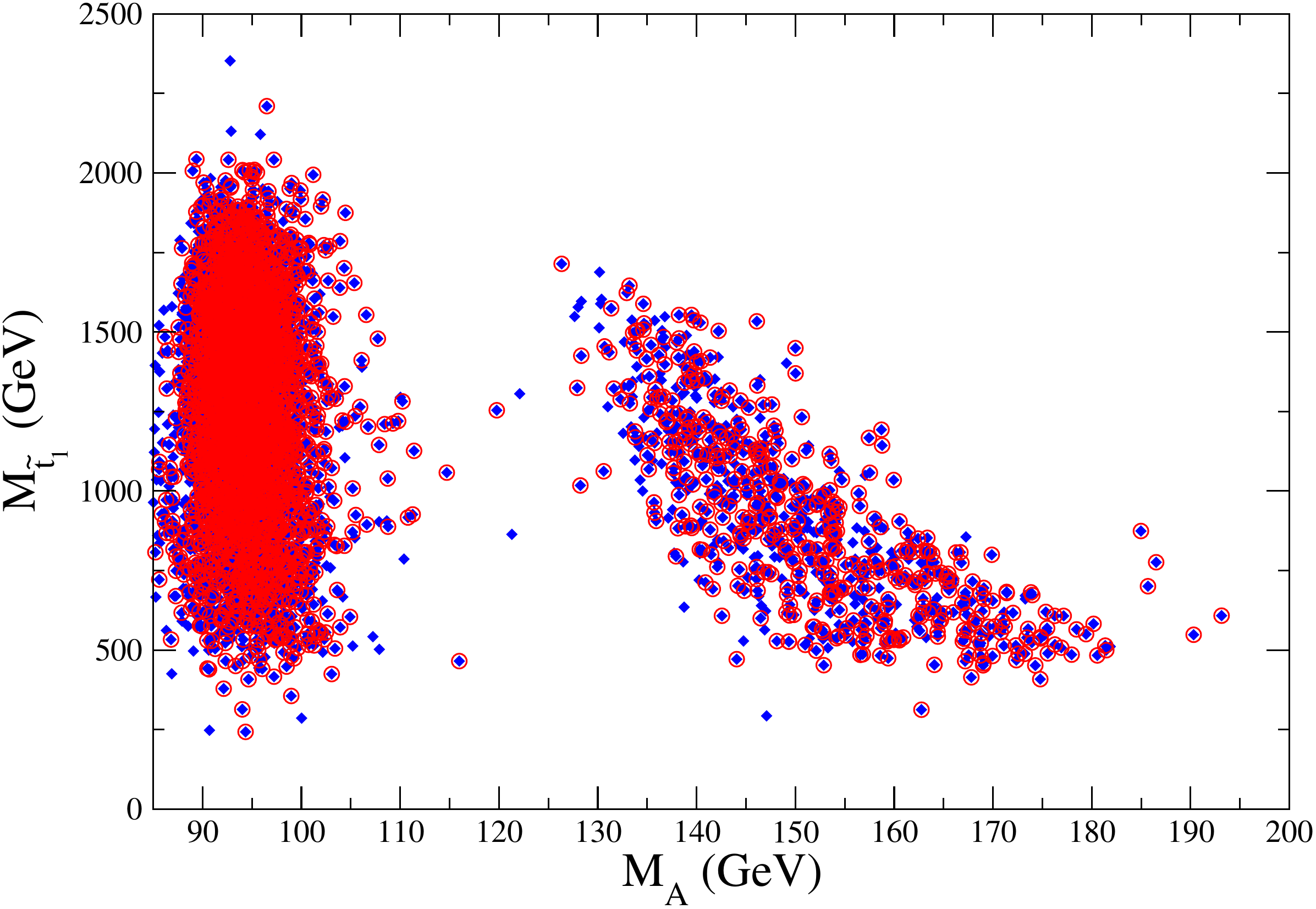}
\caption{\small
{\it
a) Scatter plot of mass of pseudoscalar Higgs boson $M_A$ vs 
the Higgsino mixing parameter $\mu$ in the 
ILLH scenario of MSSM and b) Scatter plot for 
$M_A$ vs $M_{{\tilde t}_1}$. Symbols have same meaning as in Fig.\ref{MA-tanbeta}.
}
}
\label{MA_vs_muAndStop}
\end{figure}\\
Finally, Fig.\ref{ma_mu_actual} shows the possible values of 
the higgsino mixing parameter 
$\mu$ in $M_A-\mu$ plane when all the constraints 
(Eqs.\ref{lep-lhc-masslimits} to \ref{planckdata}) are 
imposed.  For the valid $M_A$ zone of Eq.\ref{survivingMA} 
most of the scattered points correspond to values of $\mu$ 
between 3.5~TeV to 6~TeV.
Fig.\ref{ma_vs_mstop_actual} shows the spread of the lighter top-squark 
mass in $M_A-M_{{\tilde t}_1}$
 plane. For the valid $M_A$ zone 
satisfying Eq.\ref{survivingMA} we find  
the value of lighter top-squark mass to spread within 
$400 \rmGeV < M_{{\tilde t}_1}< 1.6 \rmTeV$, a significant 
part of the range may indeed be probed in the LHC. 

\subsubsection{Higgs decay channels}
\begin{figure}[!htb]
\vspace*{0.3in}
\begin{center}
\includegraphics[scale=0.4]{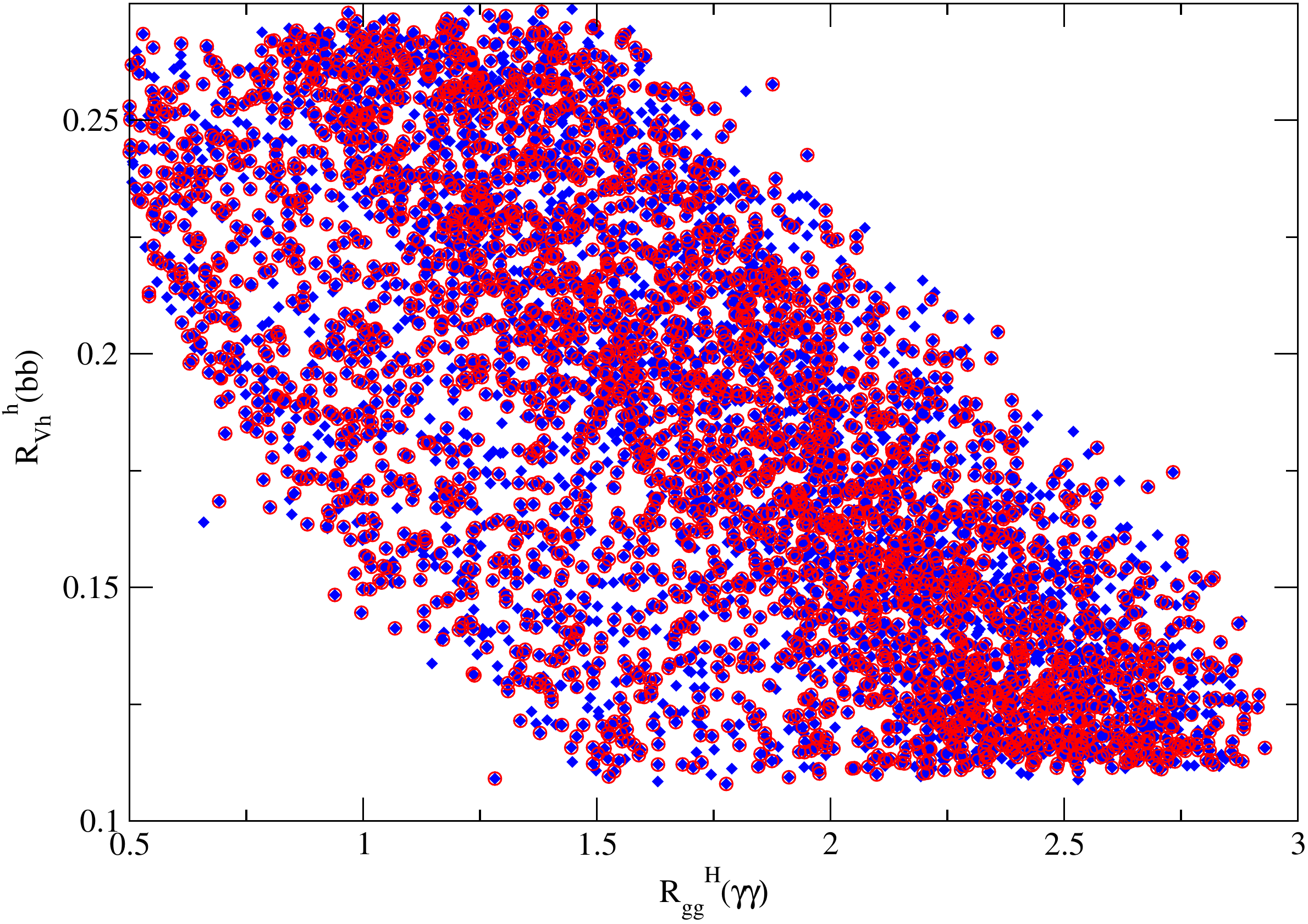}
\caption{{\small
{\it
Scattered plot of $R_{gg}^H(\gamma \gamma)$ vs $R_{Vh}^h(b \bar b)$ in the 
ILLH scenario of MSSM.
The diamond (blue) shaped points satisfy the constraints of 
Eqs.\ref{lep-lhc-masslimits} to \ref{bsmumulimit}. The 
(red) circles (enclosing diamonds) additionally satisfy the DM relic 
density constraint.
}
}
}
\label{RggcapHgamma-RVsmallhhb}
\end{center}
\end{figure}

Now we compute few Higgs decay ratios of interest in our analysis. 
The latest 
results as announced in Moriond Conferences in La Thuile (March 2013) may be  
seen in Table~\ref{HiggsDecayTableData}\cite{Ellis:2013lra,Moriond1,Moriond2,
cmsupdate2013,atlasupdate2013,tev2013}. 
\begin{table}[h]
\begin{center}
\begin{tabular}[ht]{|l|l|c|}
\hline
Higgs decay channel & Experiment  & Signal strength \\
\hline
\hline
\hsm$ \rightarrow b \bar b$ & Tevatron  &  $1.6 \pm 0.75 $ \\
\hline
\hsm$\rightarrow \tau^+ \tau^-$ & CMS & $1.1 \pm 0.4$  \\
\hline
\hsm$\rightarrow \gamma \gamma $ & ATLAS & $1.65{ }^{+0.34}_{-0.30}$ \\
\hline
\hsm$\rightarrow \gamma \gamma $ & CMS & $0.78{ }^{+0.28}_{-0.26}$ \\
\hline
\hsm$\rightarrow W W^*$ & ATLAS  & $1.01 \pm 0.31 $ \\
\hline
\hsm$\rightarrow Z Z^*$ & ATLAS  & $1.5 \pm 0.4 $ \\
\hline
\end{tabular}
\end{center}
\caption{\small {\it Higgs decay channels, Experiments and Signal strengths.}}
\label{HiggsDecayTableData}
\end{table}
Fig.\ref{RggcapHgamma-RVsmallhhb} shows the result of the computation of 
$R_{gg}^H(\gamma \gamma)$ and $R_{Vh}^h(b \bar b)$ for the given parameter 
space of the ILLH scenario of MSSM. Considering $1\sigma$ limit 
of the CMS result of \hsm$\rightarrow \gamma \gamma$ from 
Table~\ref{HiggsDecayTableData} we use a conservative lower bound for 
$R_{gg}^H(\gamma \gamma)$, namely $R_{gg}^H(\gamma \gamma)>0.5$. 
The spread of $R_{Vh}^h(b \bar b)$ is consistent with the
LEP excess\cite{Barate:2003sz} of Eq.\ref{LEPcriterion}. 
The diamond (blue) shaped 
points refer to satisfying all the relevant constraints (as those from 
Eqs.\ref{lep-lhc-masslimits} to \ref{bsmumulimit}) except the DM constraint.  
The points 
denoted by red circles additionally satisfy the DM constraint. 
We note that 
the smallness of the value of $R_{Vh}^h(b \bar b)$ is consistent with the  
non-observability of $h$ in Tevatron or in LHC\cite{Belanger:2012tt} until 
now. 

\noindent
Fig.\ref{RggcapHgamma-RVcapHHb} shows the scatter plot of 
$R_{gg}^H(\gamma \gamma)$ vs $R_{VH}^H(b \bar b)$, whereas 
Fig.\ref{RggcapHgamma-RVcapHHtau} shows the same for 
$R_{gg}^H(\gamma \gamma)$ vs $R_{VH}^H(\tau^+ \tau^-)$
in the 
ILLH scenario of MSSM. 
Symbols have the same meaning as in 
Fig.\ref{RggcapHgamma-RVsmallhhb}. 
 We note that QCD and SUSY QCD 
corrections to $m_b$ play important roles in modifying the 
total decay width as well as relevant branching ratios of 
$H$-boson\cite{Bechtle:2012jw,Hagiwara:2012mga,Carena:2013iba}.

\begin{figure}[!htb]
\vspace*{0.3in}
\mygraph{RggcapHgamma-RVcapHHb}{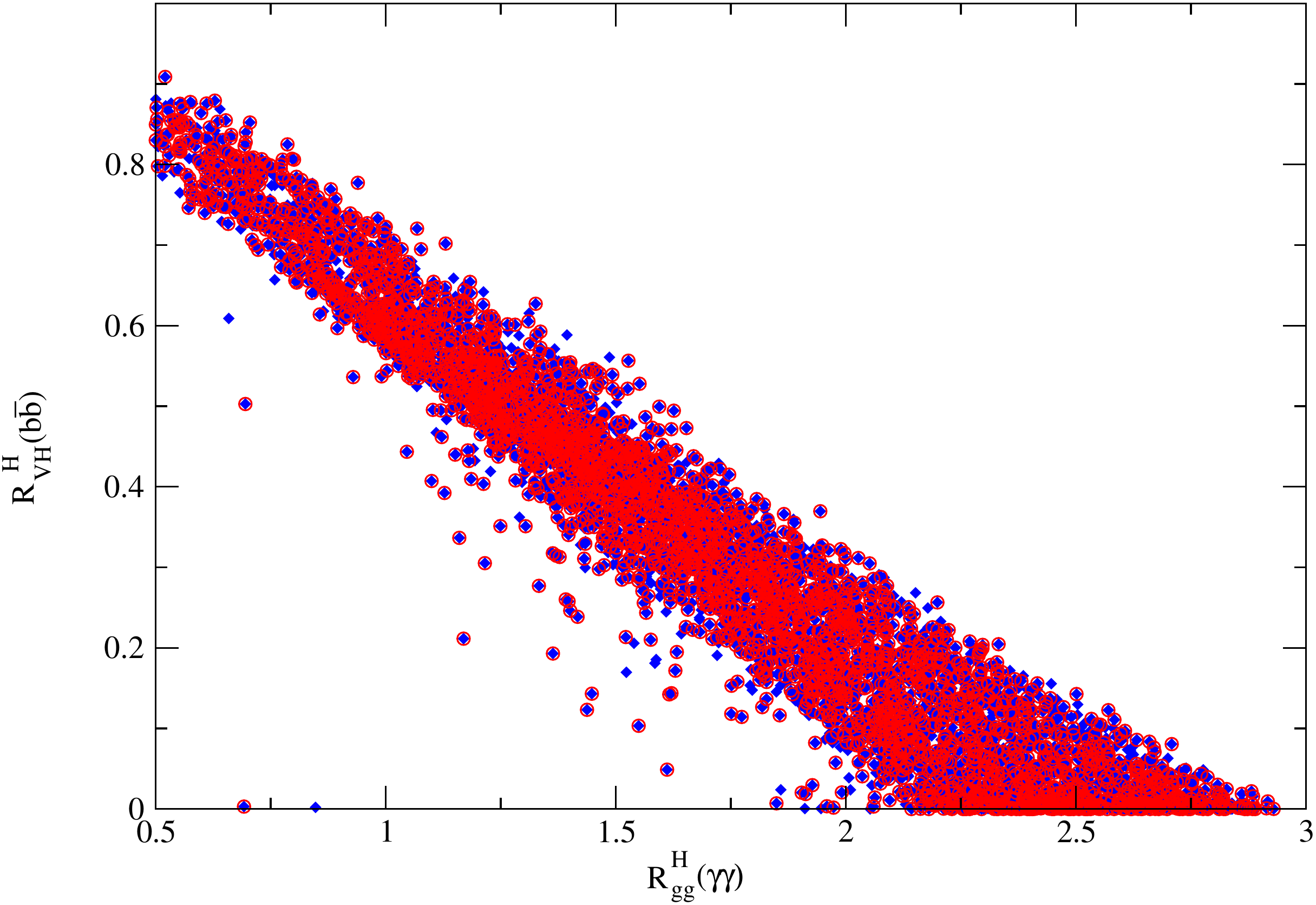}
\hspace*{0.5in}
\mygraph{RggcapHgamma-RVcapHHtau}{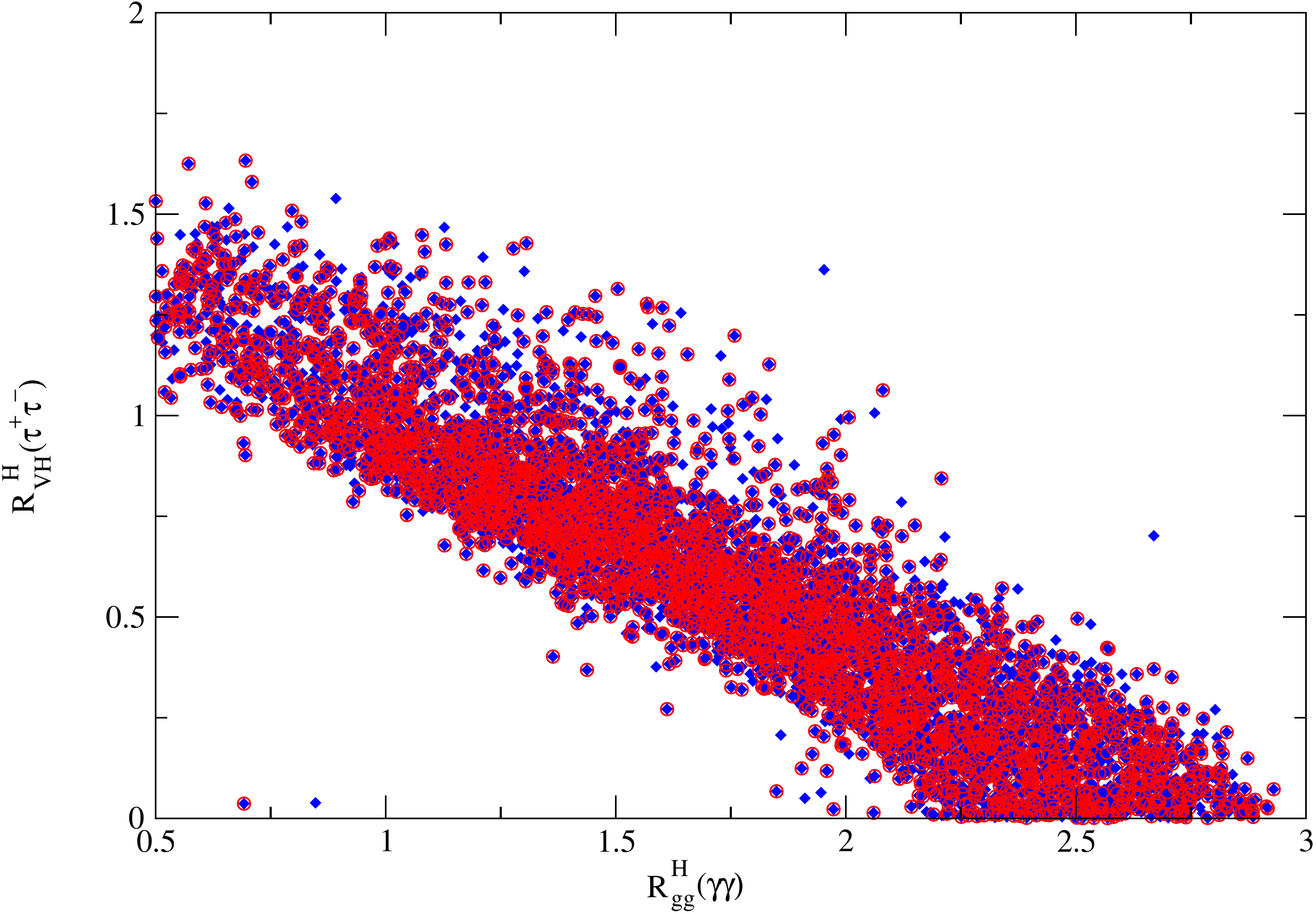}
\caption{\small
{\it
a)  Scatter plot of 
$R_{gg}^H(\gamma \gamma)$ vs $R_{VH}^H(b \bar b)$ in the 
ILLH scenario of MSSM. Symbols have the same meaning as in 
Fig.\ref{RggcapHgamma-RVsmallhhb}.
b) Scatter plot of $R_{gg}^H(\gamma \gamma)$ vs $R_{VH}^H(\tau^+ \tau^-)$ 
in the ILLH scenario of MSSM. 
Symbols have the same meaning as in 
Fig.\ref{RggcapHgamma-RVsmallhhb}.
}
}
\label{RggcapHgamma-HbbHtautau}
\end{figure}

\subsubsection{Dark matter direct detection}
We will briefly discuss now the cold dark matter relic density constraint.
We only exclude the parameter points that lead to over-abundant 
relic densities, thereby we also include the cases where the LSP  
may be a sub-dominant component of dark matter.    
The parameter ranges of Eq.\ref{parameterRanges} selected in 
our analysis are such that the principal mechanism to satisfy the relic 
density is ${\tilde \chi}_1^0$-${\tilde \chi}_1^\pm$ coannihilation where 
${\tilde \chi}_1^0$ is almost a pure bino and ${\tilde \chi}_1^\pm$ is almost 
a pure wino. This happens for more than 99\% of parameter points 
satisfying the upper limit of Eq.\ref{planckdata}. Very few parameter points 
(less than 1\%) are associated with ${\tilde \chi}_1^0$-${\tilde t}_1$ 
coannihilation so as to reduce the 
relic density for a typically bino dominated LSP to an acceptable value.       
We note that our parameter ranges as mentioned in Eq.\ref{parameterRanges} 
have heavy sleptons that naturally would not undergo any coannihilation 
with LSPs. We explore the detection prospect of the LSP in Fig.\ref{no_scaling_sig_mchi} 
by computing the spin-independent direct detection ${\tilde \chi}_1^0-p$ 
cross-section using micrOMEGAs (version 2.4.5)\cite{Belanger:2008sj,Belanger:2005kh,Belanger:2006is} with default input values for direct detection. 
A considerable region above the 
solid (black) line is discarded via XENON100 data\cite{Aprile:2012nq}.
Here the diamond (blue) marked points refer to parameter points 
that satisfy all the relevant constraints of  
Eqs.\ref{lep-lhc-masslimits} to \ref{bsmumulimit} except that they may or may 
not satisfy the DM relic density constraint. Circles (red) refer to 
satisfying only the upper limit of DM relic density of Eq.\ref{planckdata} in 
addition to satisfying the constraints same as those of the 
diamond (blue) marked points.  
We extend our analysis in 
Fig.\ref{with_scaling_sig_mchi} by computing the scaled 
cross-section ($\zeta \sigma^{SI}_{p{\tilde \chi}_1^0}$) since 
most of the parameter points correspond to under-abundant 
relic densities. 
Here, $\zeta={\rm min}\{1,\Omega_{{\widetilde \chi}_1^0} h^2/{(\Omega_{CDM} h^2)}_{\rm min}\}$\cite{BottinoRescaled},
 where ${(\Omega_{CDM} h^2)}_{\rm min}$ refers to the lower limit of 
Eq.\ref{planckdata}. Clearly a significant amount of parameter 
space will be 
probed in future direct-detection experiment XENON-1T\cite{Aprile:2012zx}.
In Table~\ref{tab:bp0} we show an example benchmark point for the 
ILLH scenario in MSSM in relation to 
Eq.\ref{parameterRanges}.

\begin{table}[htb]
\centering
\small
\begin{tabular}{ccccccccc}
\hline
$M_t$ & $M_A$ & $\tan\beta$ & $\mu$ & $M_1$ & $M_2$ & $M_3$ & $A_t$ & $A_b$ \\[1mm]
\hline
173.6 & 167.5 & 5.0 & 5429.8 & 527.9 & 119.2 & 1416.6 & 5729.2  & -217.1 \\ 
\hline\hline
$A_{\tau}$ &  $M_{\tilde{q}_{3L}}$ & $M_{\tilde{t}_R}$ & $M_{\tilde{b}_R}$ & $M_h$ & $M_H$ & $M(H^\pm)$ 
& $M_{\tilde{t}_1}$ & $M_{\tilde{b}_1}$ \\[1mm]
\hline
-115.2 & 1712.6 & 1602.2 & 426.7 & 97.7 & 125.1 & 182.1 & 999.2 & 539.1 \\ 
\hline\hline
$M_{\tilde{g}}$ &  BR($B_s \to \mu^+ \mu^-)$ & BR($b \to s \gamma)$ & $\Omega h^2$ &  
$\zeta \sigma^{SI}_{(p-\chi)}$  &  &  &  \\[1mm]
\hline
1608.9 & 2.8$\times10^{-9}$ & 3.8$\times10^{-4}$ & 4.5$\times10^{-4}$ & 5.5$\times10^{-11}$  &  &  &  \\
\hline
\end{tabular}
\def\baselinestretch{1.0}
\caption{\small {\it{ An example benchmark point for the ILLH scenario in 
MSSM in relation to Eq.\ref{parameterRanges}. 
All masses are in units of GeV.}}}
\def\baselinestretch{1.0}
\label{tab:bp0}
\end{table}
\begin{figure}[!htb]
\vspace*{0.3in}
\mygraph{no_scaling_sig_mchi}{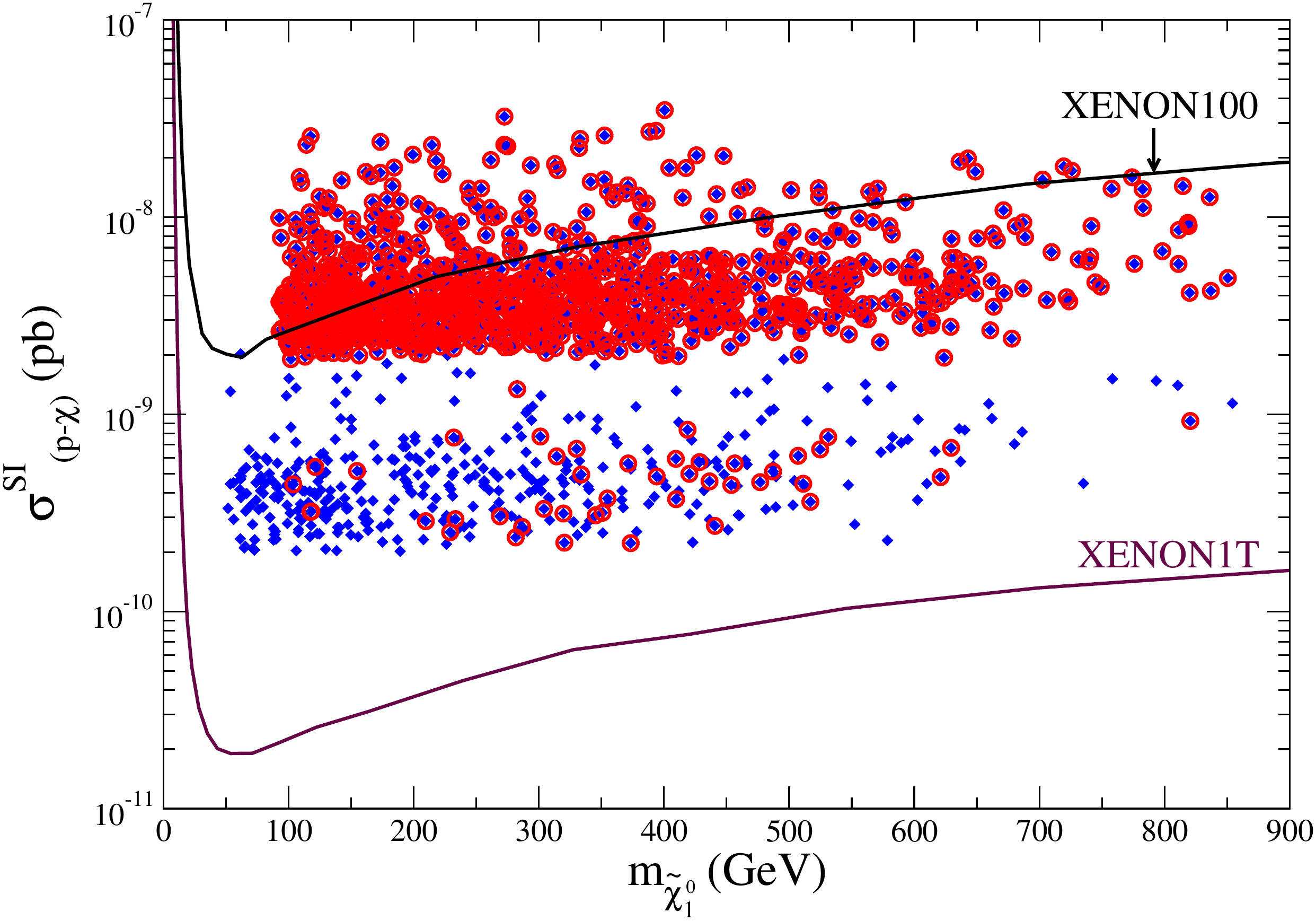}
\hspace*{0.5in}
\mygraph{with_scaling_sig_mchi}{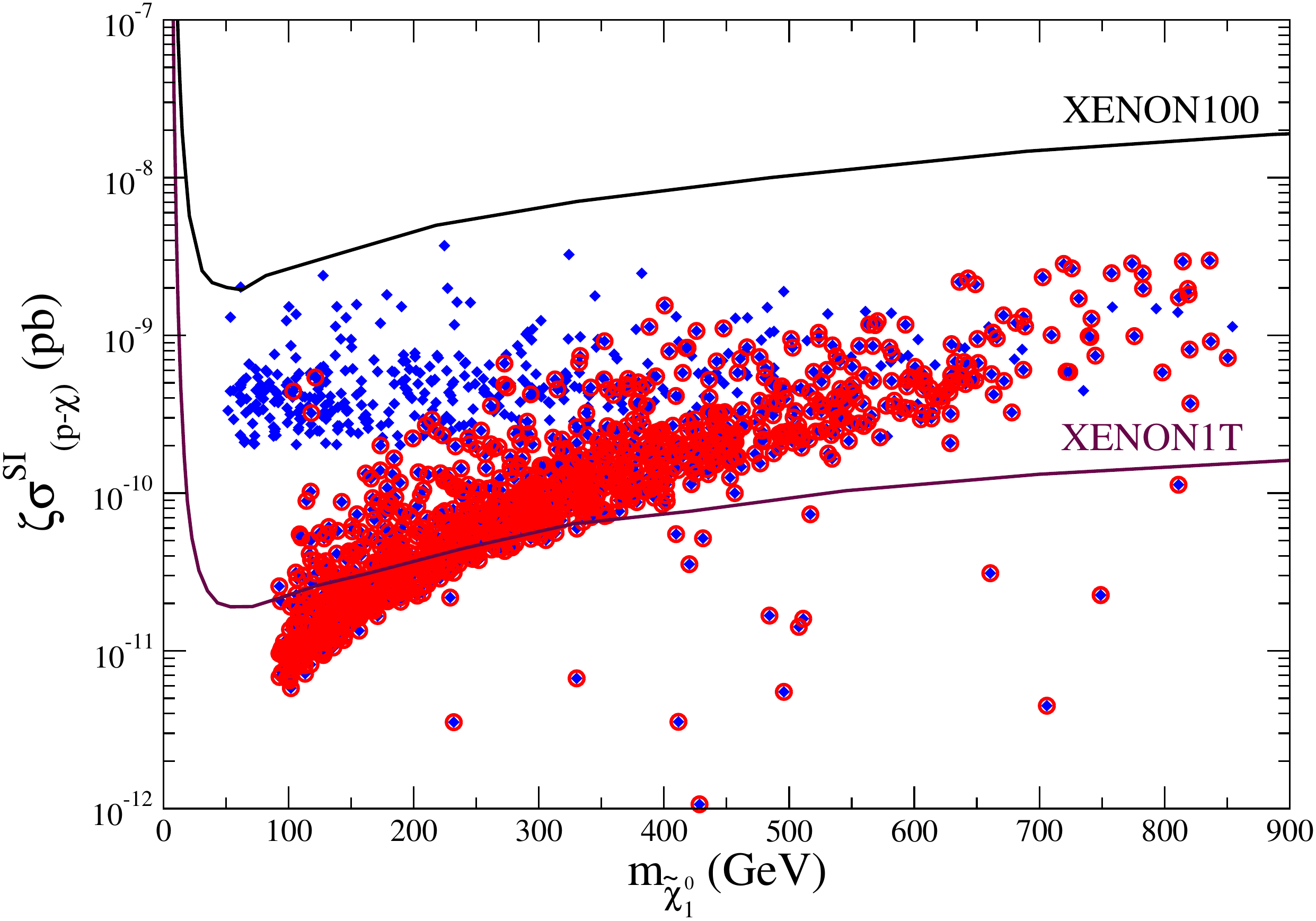}
\caption{\small 
{\it
a) Scatter plot of $m_{{\tilde \chi}_1^0}$ vs 
the spin-independent proton-${\tilde \chi}_1^0$ direct detection 
scattering cross-section $\sigma^{SI}$ in the 
ILLH scenario of MSSM. The XENON100 and future 
experiment of XENON-1T constraints are shown as solid lines. 
Further details are mentioned in the text. 
 b) Same as (a) except that $\sigma^{SI}$ is rescaled to  
$\zeta \sigma^{SI}$, where $\zeta={\rm min}\{1,\Omega_{{\widetilde \chi}_1^0} h^2/{(\Omega_{CDM} h^2)}_{\rm min}\}$. 
}
}
\label{sig_mchi}
\end{figure}

In the next subsection we would like to extend our studies of the 
ILLH scenario to Next to Minimal Supersymmetry 
Model (NMSSM).
However, before we move on to NMSSM Higgs sector, we would like to 
highlight the main issues of our analysis in MSSM:
\begin{itemize}
\item 
We have found that the constraints from CMS and ATLAS for the 
ILLH scenario where one can possibly have both 98 GeV 
as well as 125 GeV for the two Higgs boson masses  
restrict $\tan\beta$, $M_A$ and $M_{H^\pm}$ 
to be within the following ranges: $3< \tan\beta <5.5$, 
      $ 130~{\rm GeV} < M_A < 200~{\rm  GeV} $ and  
       $ 150~{\rm GeV} < M_{H^+} < 200~{\rm  GeV} $.
\item Parameters related to radiative corrections to the Higgs sector
     do have significant impact in our analysis. Additionally, 
constraint from 
$Br(b \rightarrow s \gamma)$ plays an 
effective role to discard a large 
region of parameter space.  Particularly this limits $A_t$ to have 
only the positive branch and constrains $\mu $ to assume 
values between $3.5 - 6$~TeV for all the allowed points 
within the valid $M_A$ zone satisfying Eq.\ref{survivingMA}.
\end{itemize}
      
\subsection{NMSSM}
One can very easily extend MSSM to NMSSM by simply adding 
a gauge singlet superfield $\hat S$ that
couples with the Higgs doublets via an interaction
$\lambda {\hat S} {\hat H}_u {\hat H}_d$ in the superpotential apart from a
self interaction term for $\hat S$ namely,
$\frac{\kappa}{3} {\hat S}^3$\cite{NMSSMreviewetc}.
$\lambda$ is chosen so that the 
above interaction terms remains perturbative up to the unification scale
($M_G \sim 2 \times 10^{16}$~GeV) depending on $\kappa$ and $\tan\beta$.
For SUSY breaking, in addition to the MSSM soft-SUSY breaking terms,
one considers
scalar mass term corresponding to the singlet superfield $\hat S$, trilinear
coupling terms $\lambda A_\lambda H_u H_d S$ and $\frac{1}{3}\kappa A_\kappa S^3$ 
along with their hermitian conjugates.

    In NMSSM one expands around the vacuum with non-vanishing vevs of the
neutral CP-even components of $H_u$, $H_d$ and $S$. The scalar component of
superfield $\hat S$ can mix with the neutral scalar
components of ${\hat H}_u$ and ${\hat H}_d$. This leads to
three CP-even Higgs bosons $H_i$, $i=1,2,3$ and two CP-odd neutral
Higgs bosons $A_i$, $i=1,2$.  Similarly, the fermionic superpartner
of $\hat S$ mixes with the neutral fermionic superpartners of ${\hat H}_u$
and ${\hat H}_d$ along with the neutral electroweak gauginos leading to
five neutralinos.  

We consider a semi-constrained version of NMSSM \cite{NMSSMreviewetc}
which is characterized by the following parameters to be given at
$M_{\rm GUT}$:
\begin{itemize}
\item Universal gaugino mass parameter $M_{1/2}$, 
\item Common scalar (sfermion) mass parameter $m_0$ except for the
Higgs scalars.  The Higgs soft mass
terms $m_{H_u}^2$, $m_{H_d}^2$ and $m_{S}^2$ are taken
as non-universal, and 
\item Common trilinear coupling for top, bottom and tau:
$A_t=A_b=A_\tau \equiv A_0$. The trilinear
couplings $A_{\lambda}$, $A_{\kappa}$ may differ from $A_0$.
\end{itemize}
Thus, the complete parameter space is defined by
\beq \label{eq: ---}
\lambda\ , \ \kappa\ , \ \tan\beta\ ,\
\mu_{\mathrm{eff}}\ , \ A_{\lambda} \ , \ A_{\kappa} \ , \ A_0 \ , \
M_{1\over2}\ ,  {\rm and } \ m_0. \
\eeq
Here, $\mu_{\mathrm{eff}}=\lambda<S>$ where $<S>$ denotes the vev of the 
scalar part of the singlet superfield.
We scan these parameters in the following ranges:
\begin{eqnarray}
0.1 < \lambda < 0.7,~~~~~&\ \  0.05 < \kappa < 0.5, \nonumber \\
1 < \tan\beta < 10,~~~~~&\ \   0.1~ {\rm TeV} <\mu_{eff} < 0.5~ {\rm TeV},  \nonumber \\
0.1~ {\rm TeV} < m_0 < 3~ {\rm TeV},&\ \  0.1~ {\rm TeV} < M_{1 \over 2} < 3~ {\rm TeV}, \nonumber \\
-1~ {\rm TeV} < A_{\lambda} < 1~ {\rm TeV},&\ \  -1~ {\rm TeV} < A_{\kappa} < 1~ {\rm TeV} , \nonumber \\
-6~ {\rm TeV} < A_{\rm 0} < 6~ {\rm TeV}.
\end{eqnarray}
The ranges of $\lambda$, $\kappa$, $\mu_{eff}$, $A_\lambda$ and $A_\kappa$ are chosen so that
the CP-even and the CP-odd Higgs boson masses may assume a wide range of values.
The random scan of the above NMSSM parameter space
to calculate the Higgs and the SUSY particle spectrum
along with various couplings, decay widths and branching ratios,
is performed
by using the code
NMSSMTools\,3.2.4~\cite{Ellwanger:2004xm,Ellwanger:2005dv}.
The dark matter relic density and direct detection cross-section
of the lightest neutralino
$\chi^0_1$  have been computed by
using micrOMEGAs (version 2.4.5)~\cite{Belanger:2008sj,Belanger:2005kh,Belanger:2006is}
as implemented in the NMSSMTools. 
Similar to the analysis with MSSM, we impose the Higgs mass 
limits for $m_{H_1}$ and $m_{H_2}$ from Eq.\ref{lep-lhc-masslimits} where 
$H_1$ here refers to the lighter CP-even Higgs boson. We also use 
all other constraints similar to what is done for MSSM.
\begin{figure}[!t]\centering
\vspace*{-1cm}
\includegraphics[angle=0,width=100mm]{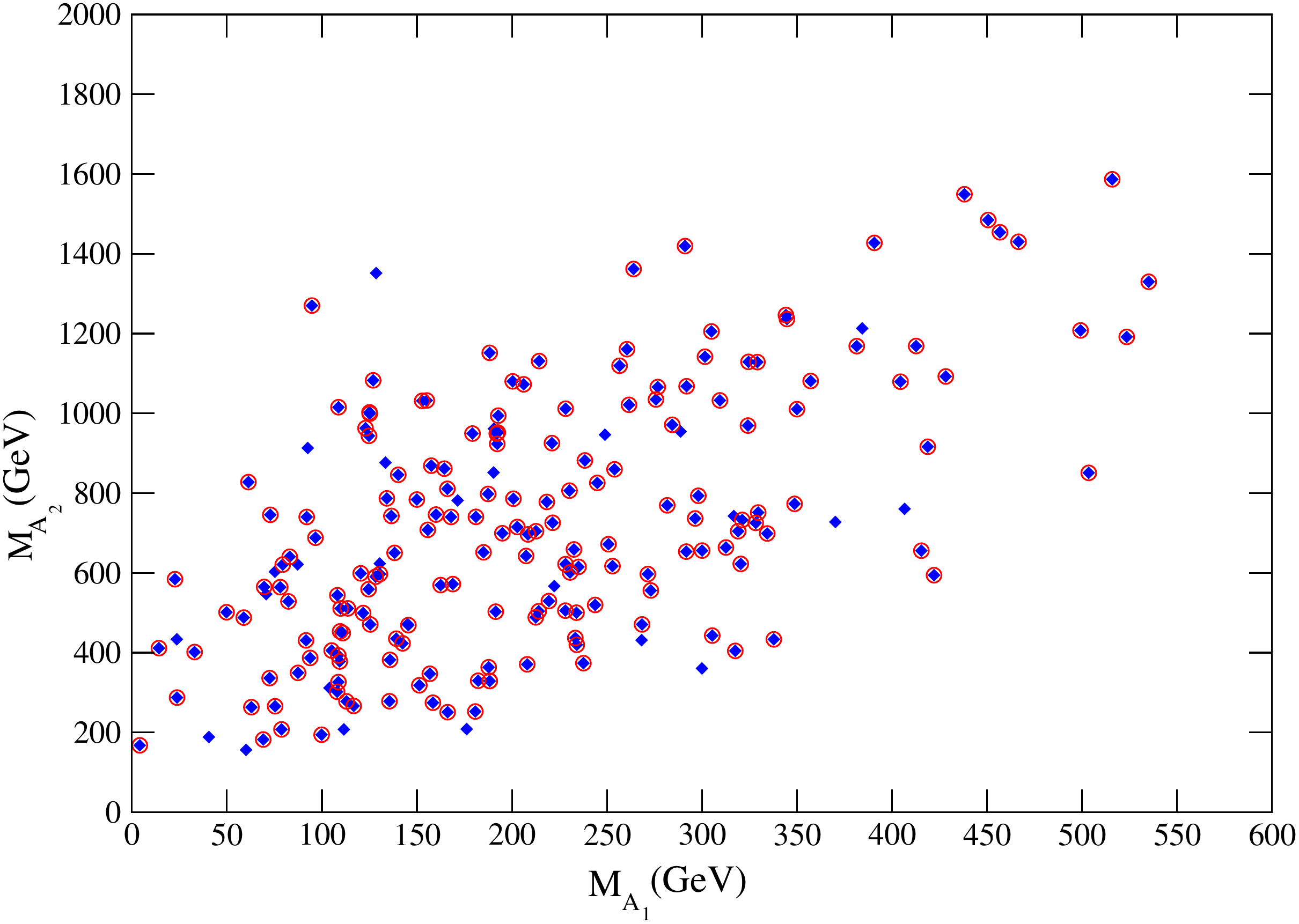}
\caption{\small 
{\it
Scatter plot in $M_{A_1} - M_{A_2} $ plane. Constraints imposed
are same as MSSM and symbols have the same meaning as in Fig.\ref{MA-tanbeta}.
}
}
\label{nmssmfig1}
\end{figure}

Fig \ref{nmssmfig1}, shows the correlation between the pseudoscalar Higgs masses in 
$M_{A_1}-M_{A_2}$ plane. Note that, in this parameter space
of interest all other Higgs masses are almost degenerate i.e. $ M_{A_2} \sim M_{H_3} \sim M_{H^\pm}$.
Thus the above along with Fig.\ref{nmssmfig1} indicates that the masses of $A_1$, $A_2$ and $H^\pm$ bosons
may become much heavier than the corresponding Higgs bosons of MSSM.
In fact, the charged Higgs boson may even be in the TeV range
while $A_1$ can be light and this is also true for heavy $A_1$\cite{Belanger:2012tt}.
This is a clear distinction in contrast to MSSM where it is very difficult to
accommodate a $98$~GeV Higgs boson with $M_A$ larger than 200 GeV
satisfying all the present collider bounds coming from the LHC data. (see Fig.\ref{MA-tanbeta}).
In Section-\ref{mssmsubsectionforresults}, we have shown that the $98$~GeV Higgs boson scenario has been
constrained to a very narrow region in the MSSM parameter space
via $A\rightarrow \tau^+ \tau^-$
searches~\cite{CMSdataH2TauTau} in CMS and $H^\pm \rightarrow \tau\nu_\tau$ searches in ATLAS~\cite{Aad:2012tj}.
However, the situation is different in the case of NMSSM.
Here the above CMS data cannot be directly applied to
constrain the parameter space associated with the ILLH scenario
of Higgs Bosons. In NMSSM the experimental bounds on the decay of neutral 
Higgs bosons to a pair of $\tau$-lepton would not be effective enough 
to constrain the model because of the presence of
doublet-singlet mixing in the NMSSM Higgs sector. 
In the following section we will discuss the discovery potential of light 
Higgs boson at the LHC.

\section{Collider Prospects}
\label{collidersection}

In the previous two sections, we studied the possibility to accommodate 
a combination of a 98 GeV and a 125 GeV Higgs bosons 
in the non-decoupling limit of MSSM and in NMSSM. One attractive feature of the non-decoupling scenario 
of MSSM is that here all the Higgs bosons are relatively light and thus can be probed at the early run of the LHC.
Hence, non-observation of these light Higgs bosons will of course indirectly exclude the possibility of a scenario 
with a 98 GeV Higgs boson. A dedicated analysis to explore this possibility at the 8 TeV and 14 TeV LHC 
run has been performed in Ref.\cite{Christensen:2012si} where $M_A $ is 
varied between $\sim 95$ to $130~{\rm GeV}$.  It is to be noted that the above 
analysis relies on the presence of $\tau$ from the decay of A or $H^{\pm}$ bosons. 
However as seen in Fig.\ref{Mch-tanbeta} that for some region of the parameter space, $H^{\pm}$ can be heavier 
than top quark and thus charged Higgs branching to $H^+ \rightarrow  \tau^+ 
\nu_\tau$ can be negligibly small. In other words, 
the above analysis is not 
applicable for $M_H^{\pm} > M_t$ region and further study is required to exclude this region. As we found in Section-\ref{mssmsubsectionforresults} 
for MSSM results (Fig.\ref{MA-tanbeta}) the above 
range of $M_A$ considered in Ref.\cite{Christensen:2012si} is now 
disfavored via the charged Higgs search by the ATLAS Collaboration\cite{Aad:2012tj}. The ATLAS search restricts 
the allowed points in the MSSM Higgs sector to be confined in the range  
$130~{\rm GeV} < M_A < 200 $ GeV 
for $\tan\beta \sim 3 - 5.5 $. 

NMSSM differs from MSSM because of  
an additional mixing in the Higgs sector arising from the singlet scalar 
in the model. Here, the lightest neutral CP odd Higgs boson can
be light or heavy depending upon the choice of parameters 
(Fig.\ref{nmssmfig1}).  Besides, heavy scalars 
or mostly singlet scalars are also difficult to produce at the LHC. 
This implies that 
it is very hard to rule out a 98 GeV Higgs boson from indirect evidences, like 
non observation of other Higgs scalars. Hence, the exclusion/discovery of a 
98 GeV Higgs boson is completely a model dependent phenomena and we need some way 
out to exclude this possibility in a model independent manner. 
We would like to stress the fact that a statistically significant
$2.3  \sigma$ excess of events in the LEP experiments
constrains the effective
coupling ${g^{BSM}_{ZZH}}$ leading to
${g^{BSM}_{ZZH}}/{g^{SM}_{ZZH}} \simeq 0.3-0.5$. We will see  
that the above range of the ratio 
in turn controls the Higgs production cross-section at LHC.  

As mentioned previously, 
the primary Higgs boson 
production channels relevant at the LHC 
environment are gluon-gluon fusion. 
For $m_H \le $ 130-140 GeV, the Higgs boson primarily decays to b-quarks, 
tau leptons, and $WW^*/ZZ^*$. 
It is particularly difficult to detect a 98 GeV Higgs boson
produced via gluon-gluon fusion and
decaying into bottom quarks because of large QCD jet background.
Besides, if the Higgs decays to a pair
of photons, it is again very hard to distinguish it
from the continuous background due to its
heavily suppressed branching ratio to a pair of photons
($R_{gg}^{H}(\gamma \gamma) \sim 2-5\%$).
One may also produce such a light Higgs boson 
via vector boson fusion process which then decays into a pair of 
$\tau$-lepton. The signal consists of two tagged forward jets and a pair of 
$\tau$-leptons in the central rapidity region. It was shown that 
the detection of the Higgs boson via this channel is most sensitive 
for the Higgs mass in the vicinity of 130 GeV\cite{Plehn:1999nw,Green:2005ye}. 
Consequently, a VBF production mode would not be sensitive 
enough to detect a 98 GeV Higgs boson because of a large background.
Hence, we ignore this production mode in rest of
our analysis.
The next most important production mode is the Higgs-strahlung process 
$VH$ where the $H$ is produced in association with a gauge boson $W/Z$.
The Higgs-strahlung process $VH$ in a boosted regime signifies that
both the gauge bosons would have large transverse momenta.  
However, this boosted regime corresponds to a very small fraction
of the total cross-section 
(about 5\% for $p_T^H > $200~GeV, for $m_H=120$~GeV)\cite{Butterworth:2008iy} 
for a 14 TeV LHC run. 
On the other hand, the kinematic acceptance for this process is relatively large,
whereas the backgrounds are small. The most important feature of a boosted particle
decaying into multiple hadronic jets is that the final states remain
highly collimated because of large transverse
momenta of the parent particle. The latter appear
as a single fat jet. Hence, conventional jet finding algorithms
would not be sufficient enough to reconstruct the signal events.
This realization has led to a new technique,
called {\it{Jet Substructure}}\cite{Butterworth:2008iy} that
studies new physics signatures involving hadrons in final states.
This technique is based on the following sequential algorithm:
\begin{itemize} 
\item  Final state $b$-jets (from $ H \to b {\bar b}$ ) are formed based 
iterative jet clustering algorithm. Here we consider the Cambridge Aachen 
algorithm \cite{Dokshitzer:1997in}.   
\item Investigate subjet kinematics step-by-step. 
\item Choose the best subjets to form the fat jet mass which 
    essentially corresponds to the parent Higgs particle.
\end {itemize}
Phenomenological analysis employing state-of-the-art jet
substructure technique for highly boosted regime show
that the Higgs-strahlung process could be a very promising search
channel for a 120 GeV SM Higgs boson\cite{Butterworth:2008iy}.
The ATLAS collaboration investigated this claim using a further
realistic simulation \cite{atlaspublic}.
The ATLAS analysis performed in three sub-channels
based upon the decay of the vector boson considered the following.
\begin{enumerate}
\item Missing~transverse~momentum~ $\PMET > 30$~GeV 
and $ p_T ^{e /\mu} >30$ ~{\rm GeV} 
consistent~with~a $W$-boson with $ p_T > p_T^{\rm min} $.
\item Di-lepton invariant mass cut : $80~{\rm GeV} < m_{\ell \ell} 
< 100$ GeV, where, $\ell = e, \mu $ and $ p_T > p_T^{\rm min} $.
\item Missing~transverse~energy $ \EMET > p_T^{\rm min} $.
\end{enumerate}
\noindent
Here, $p_T^{\rm min}$ refers to the 
minimum transverse momentum of the Higgs required to call it a
{\it{Fat jet}}. As in the ATLAS physics note of Ref.\cite{atlaspublic} 
we assume $p_T^{\rm min}=$~200 GeV.  The selection criteria (1) is mostly related to the process WH when W decays to a lepton
(e/$\mu$) and a neutrino, while (2) refers to the process HZ when Z-boson 
decays into a pair of lepton (e/$\mu$). Item (3) selects the process HZ 
when Z-boson decays invisibly to a pair of neutrinos.
Contribution from HW process may come for the sub-channel (3) when the lepton from W
is outside the acceptance domain. We note that, (2) is a very clean signal but has a low cross-section while
(1) and (3) have relatively higher signal cross-sections but these are
mostly overshadowed by continuous $t \bar{t}$ background.
Using the prescribed procedure of ATLAS simulation, we have reproduced
expected number of events with 30 ${\rm fb}^{-1}$ data for all the
three signal subprocesses as presented in the Table~1 of
the above-mentioned ATLAS physics note\cite{atlaspublic}.
We have used Pythia6 (version 6.4.24)\cite{pythia6}
for the generation of signal events and the
package FASTJET (version 2.4.3)\cite{fastjet2} for 
reconstruction of jets and implementation of the jet substructure analysis
for reconstruction of the boosted Higgs. 
We present our final results in Table~\ref{tab:sig1}, where the background
event numbers are appropriately 
scaled by using corresponding numbers given in the above 
ATLAS note.
In order to satisfy the existing LEP bounds
for the 98 GeV Higgs boson, we further assume a 20\% LEP excess
around the mass window 92 - 108 GeV. The expected number of events
for signal and the various backgrounds
for 300 ${\rm fb}^{-1}$ of collected data is listed in
Table~\ref{tab:sig1}. The resulting
statistical significance is presented in the fourth column of
Table~\ref{tab:sig1} considering the aforesaid mass window,
whereas the last column represents the combined significance
we obtained by adding the same for each channel in quadrature. We find 
that one can marginally exclude the presence of 98 GeV 
Higgs boson using 300 fb$^{-1}$ of data at the 14 TeV LHC assuming only statistical 
uncertainty. However, at these high luminosity the pile ups, 
multiple interactions play a crucial role. Consequently 
estimation of signal significance would be limited due 
to large systematic uncertainties in the background calculations. 
 On the other hand, backgrounds and selection cuts used in our analysis 
are taken from the above-mentioned ATLAS  note\cite{atlaspublic}. These are 
not optimized for the 98 GeV Higgs boson, thus requiring 
further realistic analysis. From the above discussion one can infer that the 
model independent exclusion of a 98 GeV Higgs boson may not be possible 
even at the high luminosity run of LHC. However, if we are fortunate enough, 
we can even discover this particle via other processes.
\begin{table}[t]\centering
\begin{tabular}{ccccc}
\hline
Process & Signal (S) & Background (B) & Significance ($ S \over \sqrt{B}$)
& Combined Significance \\
\hline
$ \ell \nu b \bar{b}$ & 35.6  & 417 & 1.7 & \\
$ \ell^+ \ell^- b \bar{b}$ & 11.8 & 160.5 & 0.9 & 2.5 \\
$ \EMET b \bar{b}$ & 54.6 & 1136 & 1.6 &   \\
\hline
\end{tabular}
\caption{ \small
{\it
Expected number of events at 14 TeV LHC run 
with 300 ${\rm fb}^{-1}$ of integrated luminosity within a mass window
92 - 108 GeV, based on LO cross sections, for the individual signal and the combined background
processes assuming 20\% LEP excess in this region of interest.
The combined significance is obtained by adding significances in quadrature.}
}
\label{tab:sig1}
\end{table}
 Let us survey the discovery potential of another production mode: the associated 
production of the 98 GeV 
Higgs with top quarks when the Higgs decays to a pair of bottom quark 
($pp \rightarrow t \bar t H$, $H \rightarrow b \bar b$).
The production cross-section in this channel  for a $\sim$ 100 GeV 
Standard Model Higgs is $\sim$ 1 pb at the 14 TeV run of 
LHC\cite{ttbarh}. In a boosted regime, the decay products of both the top quarks 
and the Higgs would be highly collimated
and jet substructure algorithm can be a very useful tool to tag top quarks and Higgs. 
Instead of reconstructing the individual 
top decay products, in {\it top tagging} which is a technique to identify boosted 
hadronic top quarks, one uses a jet algorithm and performs a subjet analysis to 
reconstruct the top quark mass. We would like to remind our readers that production cross section 
of this channel is highly parameter space dependent. Performing a random scan of the MSSM parameter space, we obtain
$R_{pp}^{h}(b\bar{b})$ to be within 20 to 60\% where 
$h$ refers to the 98~GeV Higgs boson of MSSM.
However, with a conservative standpoint we choose 
$R_{pp}^{h}(b\bar{b})= 0.2$.  
We do not perform any detailed analysis in this mode. Here we
refer the results of an analysis already performed in this direction
for a $\sim$ 115 GeV Standard Model Higgs boson at 14 TeV LHC \cite{Plehn:2009rk}. 
While translating the results 
of Ref.\cite{Plehn:2009rk} for our choice of Higgs mass, we expect an enhancement of $60\%$ and $20\%$ 
in the Higgs production rate and the background estimation respectively. Both these enhancements simply come 
from the difference in the choice of Higgs boson mass considered in Ref.\cite{Plehn:2009rk} and ours. 
Hence, in our analysis we scale the signal and background by 1.6 and 1.2 respectively for a 98 GeV Higgs
boson decaying to a pair of bottom quarks. It turns out that for an integrated luminosity of 300 ${\rm fb}^{-1}$ and 
a 98 GeV Higgs boson, with two tagged b-jets the statistical significance of this 
channel is $\sim$ 3.1$\sigma$. On the other hand, 
for a three b-tag sample the significance is 
$\sim$ 2.6$\sigma$.  Again, we remind
that the systematic uncertainty has not been taken into account.

Apart from these direct production mechanisms, the 
98 GeV Higgs boson may be produced from the cascade decay of SUSY particles. 
 From our scan of MSSM and semi-constrained NMSSM it comes out that squarks/gluinos are sufficiently heavy leading to very small 
strong production cross section. However, in a more general framework squark/gluinos can be light,
 just above the LHC specified limits. In that case, the decay of gauginos can produce highly boosted Higgs bosons 
which may be probed using the jet substructure technique (for details see\cite{Butterworth:2007ke,Kribs:2010hp,Stal:2011cz,
Bhattacherjee:2012bu} and references therein). Besides, a 98 GeV Higgs production from the decay of heavy Higgses may 
play an important role at the LHC\cite{King:2012tr,Kang:2013rj}. 

The proposed $e^+ e^-$ international linear collider 
(ILC) is an ideal machine to study this ILLH scenario 
due to its clean environment and relatively less background contamination.
The possibility to verify this ILLH scenario in the context of NMSSM 
in the Linear colliders have already been discussed in Ref.\cite{Belanger:2012tt}. Here, 
 the authors studied all the possible production/decay modes of NMSSM Higgs bosons in the context 
of Linear collider, Photon collider as well as Muon collider. 
However, our goal is to probe this scenario in a model independent way. 
Here we consider ILC with $\sqrt{s}$= 250 GeV and integrated luminosity =100  
${\rm fb}^{-1}$. The golden channel for the Higgs production in the context of ILC is $e^+ e^- \to ZH $, 
where, $Z$ can decay both leptonically or hadronically.
The final state would involve jets and/or leptons depending upon the decay of
the $Z$-boson while we assume Higgs always decays into a pair of bottom quark.
We use MadGraph5\cite{Alwall:2011uj} to estimate the cross-section
for both the signal and the SM background for $100 {\rm fb}^{-1}$ luminosity.
The production cross-section of a 98 GeV Standard Model Higgs boson at the ILC 
with 250 GeV center of mass energy is about 350 ${\rm fb}$. The most dominant background 
comes from the $Z$-pair production which is about 1.1 pb.  In order to satisfy the LEP bounds, 
we assume a $20\% $ LEP excess for the 98 GeV Higgs boson. 
We further assume $60\% $ $b$-tagging efficiency while calculating the statistical 
significance $(S/\sqrt{B})$. We find that a 98 GeV Higgs boson can be easily 
discovered/excluded at the 250~GeV ILC with a 100 ${\rm fb}^{-1}$ luminosity 
which is easily achievable within the first few years of its run.

\section{Conclusion}
\label{conclusionsection}
To summarize, we have studied the possibility that both the LEP
excess in $b\bar{b}$
final state with a 98 GeV Higgs boson and the LHC
di-photon signal for a 125 GeV
Higgs-like object can be simultaneously explained
in the general MSSM framework. It turned out that the MSSM
parameter space where such scenario is valid is also 
consistent with several low energy constraints obtained from 
the heavy flavour sector, the cold dark matter 
constraints obtained from the recent PLANCK collaboration, 
limits from XENON100 experiment on the dark matter direct detection 
cross-section. In our MSSM Higgs parameter space 
scan we have also implemented the recent 
results on the heavy MSSM Higgs boson search via 
$H/A \to \tau^+ \tau^- $ mode from CMS and ATLAS collaboration 
and the search of charged Higgs boson from ATLAS via $H^+ \to \tau^+ \nu_\tau $
mode. After all these constraints, we have finally obtained the 
following 
allowed ranges of $M_A$ and $M_{H^\pm }$: 
       $ 130~{\rm GeV} < M_A < 200~{\rm  GeV} $ and  
       $ 150~{\rm GeV} < M_{H^\pm} < 200~{\rm  GeV} $.
       We have also shown that for the above ranges of $M_A$ and 
$M_{H^\pm}$, $\tan\beta$ is confined 
      in a very narrow window $\sim 3- 5.5$ from the Higgs searches
      at the LEP and the LHC. 

We have also pointed out that parameters related to radiative corrections 
to the Higgs spectrum have significant impact on our whole analysis. 
Additionally, 
$Br(b \rightarrow s \gamma)$ limits discard a large 
region of parameter space and this we tried to explain in a 
reasonable detail.  Particularly, this causes 
$\mu $ to vary between $3.5 - 6$~TeV for all the 
allowed points. 

We have also studied the correlation between different decay 
modes $(H \to \gamma \gamma, ~b {\bar b}, ~\tau {\bar \tau}) $
of the heavy Higgs boson of MSSM ($M_H \sim 125 $ GeV) and our results showed
that these rates are consistent within $ 1\sigma$  of 
the latest data on the Higgs signal from 7 TeV and 8 TeV  
LHC run. Moreover, we have also discussed briefly 
the prospect of the ILLH scenario 
in the context of NMSSM. Indirect exclusion of a 98 GeV
Higgs boson in NMSSM is possible though it depends on the choice 
of the model parameters and the particle spectrum. 

We studied the possibility of observing the 98 GeV Higgs boson at the
14 TeV run of LHC. We focused on the 98 GeV Higgs production via 
Higgs-strahlung: $ p p \to VH $, followed by $ H \to b {\bar b}$ decay. 
Because of the large $p_T$ of such Higgs boson at the 14 TeV LHC, 
the decay products of the Higgs boson will form a highly collimated jet
and we performed our analysis using the technique of jet substructure. 
Our analysis showed that the statistical significance for 
observing the above 
Higgs boson signal at the 14 TeV LHC run with 300 ${\rm fb }^{-1}$ 
luminosity is at most $2.5\sigma $. 
Moreover, when the Higgs boson is produced in association with top quarks 
one can achieve a 3.1$\sigma$ (2.6$\sigma$) level of statistical significance 
with two (three) tagged $b$-jets.  
Here, we would like to stress
that throughout the analysis we have not taken into account 
the systematic uncertainties in the SM background estimation 
which certainly would modify the signal significance. 
We hope that the experimental collaborations would perform
more dedicated analysis in this direction using the real data of the 
14 TeV run of LHC. 
We have also attempted to explore the prospect of ILC to probe 
ILLH Higgs scenario. Even with a very conservative estimate, 
ILC would have a 
much better sensitivity  to exclude or discover the 98 GeV Higgs boson 
within a few years of its run.

\section{Acknowledgments}
B.B. and D.D. thanks Indian Association for the Cultivation of Science (IACS), 
India for hospitality when the initial part of work was done. B.B 
acknowledges the support of the World Premier International Research Center Initiative (WPI Initiative), MEXT, Japan.  
D.D. thanks Harish-Chandra Research Institute (HRI), 
Allahabad, India for hospitality in addition to the support received from 
the DFG, project no.\ PO-1337/3-1 at the Universit{\"a}t W{\"u}rzburg.
M.C. would like to thank the Council of Scientific and Industrial Research, 
Government of India for support. DKG would like to thank the 
Helsinki Institute of Physics, University of Helsinki, Finland for the 
hospitality during the final stage of this work.


\end{document}